\documentclass[usenatbib,useAMS,a4]{mn2e}
\usepackage[dv ips]{graphicx} \usepackage{natbib}
\usepackage{lastpage} \usepackage{subfigure}
\usepackage{float}
\let\oldhat\hat
\renewcommand{\vec}[1]{\mathbf{#1}}
\renewcommand{\hat}[1]{\oldhat{\mathbf{#1}}}

\title[Bayesian analysis of resolved stellar populations]{A transdimensional Bayesian method to infer the star formation history of resolved stellar populations}

\author[J. J. Walmswell, J. J. Eldridge, B. J. Brewer and C. A. Tout ]{J. J. Walmswell$^1$ \thanks{E-mail: jjw49@ast.cam.ac.uk}, J. J. Eldridge$^2$, B. J. Brewer$^3$ and C. A. Tout$^1$ \\
$^1$Institute of Astronomy, Madingley Rd, Cambridge, CB3 0HA, UK\\
$^2$Department of Physics, The University of Auckland, Private Bag 92019, Auckland, New Zealand \\
$^3$Department of Statistics, The University of Auckland, Private Bag 92019, Auckland, New Zealand}

\date{March 2013} \pagerange{\pageref{firstpage}--\pageref{lastpage}}
\begin{document}

\maketitle
\label{firstpage}

\begin{abstract}
We propose a new method to infer the star formation histories of
resolved stellar populations. With photometry one may plot observed
stars on a colour-magnitude diagram (CMD) and then compare with
synthetic CMDs representing different star formation histories. This
has been accomplished hitherto by parametrising the model star
formation history as a histogram, usually with the bin widths set by
fixed increases in the logarithm of time. A best fit is then found
with maximum likelihood methods and we consider the different means by
which a likelihood can be calculated. We then apply Bayesian methods
by parametrising the star formation history as an unknown number of
Gaussian bursts with unknown parameters. This parametrisation
automatically provides a smooth function of time. A Reversal Jump
Markov Chain Monte Carlo method is then used to find both the most
appropriate number of Gaussians, to avoid overfitting, and the
posterior probability distribution of the star formation rate. We
apply our method to artificial populations and to observed data. We
discuss the other advantages of the method: direct comparison of
different parametrisations and the ability to calculate the
probability that a given star is from a given Gaussian. This allows
the investigation of possible sub-populations.
\end{abstract}

\begin{keywords}

stars: statistics - stars: formation - methods: statistical

\end{keywords}

\section{Introduction}

One of the consequences of the advances in telescopic observing power
is a growing number of resolved stellar populations. These provide
opportunities to investigate not only the present state of galaxies
but also their pasts. If we have photometry in two or more filters
then a stellar population may be conveniently represented with a
colour-magnitude diagram (CMD). This yields immediate information
about its history. The main-sequence (MS) turn-off, for example,
indicates the brightest and hence most massive stars that are still
burning core hydrogen. Because more massive stars have shorter main
sequence lifetimes, the age of these stars, as predicted by stellar
models, equals the time since the last episode of star formation.

It is desirable to infer the complete star formation history (SFH) of
a stellar population. With detailed libraries of stellar models it is
possible to produce synthetic CMDs for star formation histories that
are complex functions of both metallicity and time. Distance,
extinction, multiplicity and the nature of the initial mass function
(IMF) may also be considered. The large array of possible parameters
means that, with the exception of especially simple systems, a
subjective comparison between observed and synthetic CMDs is
unhelpful. An automated process is required.

It was noted by \citet{1997NewA....2..397D} that the CMD of a
population with a complex star formation history may be represented by
the weighted sum of many \emph{partial} CMDs, each representing a
flat, constant star formation rate (SFR) over some small time interval
and at a given metallicity. This is essentially an IMF-weighted
isochrone that has been integrated over some small time range. The CMD
of the inferred star formation history is then the linear combination
of partial CMDs that provides the best fit to the observations. This
is for given values of the other parameters, i.e. distance,
extinction, etc. and these can then be varied to find the best overall
fit.

The term `best fit' encompasses a wide variety of concepts. One
convenient way to compare the synthetic and the observed CMDs is by
binning the data so that the observations become the number of stars
per bin. The synthetic CMD is then integrated over the same bins to
give the predicted number of stars. star formation history fitting has
thus traditionally been accomplished by optimising some fitting
parameter that compares these bins.

With small stellar populations one either has large bins that lose too
much information or small bins, most of which are empty. The problem
is worse if more than two filters are available; the vast majority of
the resulting multi-dimensional bins are then empty. This suggests an
unbinned approach and we consider different methods for analysing
unbinned data.

We also introduce a model-comparison Bayesian approach to star
formation history fitting. We model the star formation history as
being a series of Gaussian bursts, each with a time, width and
mass. We then use Markov Chain Monte Carlo (MCMC) methods to recover
the posterior probability distribution of the star formation
history. We avoid the perils of overfitting by using Reversible Jump
MCMC to make the number of bursts itself a variable and thus obtain
the posterior probability distribution for the number of Gaussian
bursts. We also show how the Reversible Jump procedure may be used to
make a direct comparison between differently-parametrised models.

If we believe that the Gaussian bursts have a physical meaning, rather
than being a parametrisation that happens to yield a smooth function
of time, then it is of interest to know the posterior distributions of
the parameters. This is made difficult by the label-switching
problem. We discuss solutions to this and recommend the use of
relabeling algorithms. We finally discuss how one can attempt to
assign the observed stars to different bursts and how that can help
identify separate populations. We apply our method to test populations
and observed data.

\section{Fitting as parameter optimisation}

We begin by outlining how several different statistics have been
applied to data to produce fitted model CMDs. We then explain the
advantages of unbinned versus binned methods and consider two
approaches by which an unbinned fitting parameter may be calculated. We
conclude by raising the pitfall of overfitting.

\subsection{Maximum likelihood with binned CMDs}

Given a synthetic CMD that has been divided up into bins, the mean
number of stars in the $i$th bin, $m_{i}$, is given by

\begin{equation}
m_i= \sum^M_j  r_j c_{ij},
\end{equation}

\noindent
where $r_{j}$ is the contribution to the star formation history from
partial CMD $j$, which has $c_{ij}$ stars in bin $i$. There are $M$
partial CMDs with $N$ bins in each. This must be compared to $n_{i}$,
its observed counterpart from a data set of $S$ stars.

The method of maximum-likelihood has been widely used to fit models to
data since it was popularised by R. A. Fisher in the early part of
the last century.  The best fit may be said to be the model that gives
the highest probability of the data given the model, the one that
maximises the likelihood. This model will be a point somewhere in the
space of the model parameters. 

With Gaussian errors ${\sigma_{mi}}$, the probability of observing
$n_{i}$ stars in bin $i$ when the model predicts $m_{i}$ is

\begin{equation}
P_i = \sqrt{\frac{1}{2 \pi \sigma_{mi}^{2}}} e^{-{\frac{(n_{i}-m_{i})^2}{2 \sigma_{mi}^2}}}.
\end{equation}

\noindent
The likelihood $\mathcal{L}$ is then the product of these individual
probabilities. They can be compared with a perfect prediction by
dividing by the probability of observing $n_{i}$ stars in bin $i$
when the model predicts $n_{i}$. Multiplying all these together
gives the likelihood ratio $\mathcal{L}$R and taking the logarithm
converts the product into a sum. It may be written as

\begin{equation}
-2 \ln \mathcal{L}{\rm R}= \sum^N_{i=1} \frac{(n_{i}-m_{i})^2}{\sigma_{mi}^2} +  \sum^N_{i=1}\ln \frac{\sigma_{mi}^2}{\sigma_{ ni}^2}. \label{Gauss}
\end{equation}

\noindent
Provided that ${ \sigma_{mi}=\sigma_{ni}}$, the second term vanishes
and ${ -2 \ln \mathcal{L}{\rm R}}$ is equal to the well-known $\chi^2$
statistic. Minimising this gives the maximum-likelihood solution,

\begin{equation}
\chi^2 = \sum^N_{i=1} \frac{(n_{ i}-m_{ i})^2}{\sigma_{ i}^{2}}.
\end{equation}

\noindent
However, minimising $\chi^2$ to find the best fit is inappropriate
when fitting CMDs.  As \citet{2002MNRAS.332...91D} observes, the
number of stars per bin given a particular mean is a Poisson
distribution.

\subsection{Minimising $\chi^2_\lambda$}

The likelihood ratio equivalent to $\chi^2$ when the data obey Poisson
statistics is ${\chi^2_\lambda}$. If there are $m_{i}$ stars in the
$i$th bin of the model CMD, compared with $n_{i}$ in the comparable
part of the observed CMD, the probability of the data being drawn from
the model is

\begin{equation}
P_i=\frac{m_{i}^{n_{i}}}{e^{m_i} n_{i} !}.
\end{equation}

\noindent
As before this may be divided by the probability that the $n_{i}$
stars would be observed when the mean is $n_{i}$. Multiplying these
together gives the likelihood ratio. The natural logarithm of this
when multiplied by -2 gives $\chi^2_\lambda$.

\begin{equation}
\chi^2_\lambda= 2 \sum_{\rm i}( m_{\rm i} - n_{\rm i} + n_{\rm i} \ln\frac{n_{\rm i}}{m_{\rm i}}).  \label{solve}
\end{equation}

\noindent
One can then apply a minimisation routine to find the maximum
likelihood star formation history. The gradients of $\chi^2_\lambda$
and its Hessian matrix are both simple analytic functions of the rates
$r_{j}$ and the partial CMDs. The latter can easily be shown to be
positive-definite and hence only one minimum for $\chi^2_\lambda$
exists. One can therefore use advanced minimisation
algorithms. \citet{2002MNRAS.332...91D} used the
Fletcher-Reeves-Polak-Ribiere (FRPR) algorithm \citep{Press}, which
proceeds by performing a series of line minimisations. His method has
been widely used since then \citep{2011ApJ...739....5W}.

\subsection{Minimising $\chi^2_\gamma$}

\citet{1999ApJ...518..380M} proposed instead a new statistic
$\chi^2_\gamma$ which both minimises properly and is more $\chi^2$-like
in form.

\begin{equation}
\chi^2_{\gamma} = \sum^{N}_{i=1} \frac{(n_{i}+\min(n_{i},1)-m_{i})^2}{n_{i}+1}
\end{equation}

It should be noted that $\chi^2_\gamma$ only in the limit of many data
points minimises to the maximum-likelihood means of Poisson data
\citep{2001NIMPA.457..384H}. This statistic has been used to deduce star
formation histories, most notably by \citet{2009AJ....138..558A} who
use a genetic algorithm to minimise it and have made their code
available on the Internet\footnote[1]{http://iac-star.iac.es/iac-pop}.

\subsection{The unbinned likelihood: Method 1}

Binning methods are problematic if too few stars have been
observed. So little information is then available that it is unwise to
throw any of it away by using large bins. However a fine binning
scheme results in many empty bins or bins containing but a single
star. A similar problem emerges if we have observations in more than
two filters. In principle this should be desirable because more data
is then available to constrain the star formation history. However the
data and the model are then more than two-dimensional and, by the
curse of dimensionality, most bins are again empty, even if the data
set is large.

The classical unbinned maximum-likelihood method
\citep{1996PhRvD..54.1194J} considers the limit of the Poisson
likelihood as the bins become very small so that each bin contains
either one star or zero stars. As before, $N$ is the number of bins
and $S$ is the number of observed stars. This definition of the
likelihood can be written as

\begin{equation}
\mathcal{L}_1 = \prod_{i=1}^{N} \frac{m_{i}^{n_{i}}}{e^{m_{i}} n_{i} !},
\end{equation}

\noindent
and separating out the product over the bins with one star and with zero
stars gives

\begin{equation}
\mathcal{L}_1   = \prod_{i=1}^{S}m_{i}e^{-m_{i}}\prod_{i=1}^{N-S}e^{-m_{i}}.
\end{equation}

\noindent
 If we combine the products of ${e^{-m_i}}$ we get

\begin{equation}
\mathcal{L}_1 = \prod_{i=1}^{S}m_{i}\prod_{i=1}^{N}e^{-m_{i}}.
\end{equation}

The Poisson mean per bin $m_{i}$, can be taken to be $p(x,y) {\rm d}x{\rm d}y$,
i.e. we introduce a probability density function $p(x,y)$ that depends
on position within the CMD. The second product is over the entire CMD. It becomes the
exponential of the sum over $m_{i}$, which in turn is equivalent to
integrating the CMD. If the $i$th star is at $x_i,y_i$ we have

\begin{equation}
\mathcal{L}_1 = \prod_{i=1}^{S} p(x_i,y_i) [({{\rm d}x{\rm d}y})^S e^{-\int p(x,y) {\rm d}x{\rm d}y}].
\end{equation}

\noindent
Providing that $p(x,y)$ has a fixed normalisation, the factor in the
square brackets will be constant. The likelihood is then proportional
to the product of the values of $p(x,y)$ at the location of the
observed stars.

\begin{equation}
\mathcal{L}_1 \propto \prod_{i=1}^{S} p(x_i,y_i).
\end{equation}

The obvious normalisation of $p(x,y)$ is so that it integrates to
one. For consistency one should perform the integration over a fixed
area in apparent magnitude and uncorrected colour, within which the
data points are fixed locations. It is thus aways the probability
density function of observing a star given the model. Varying the star
formation history, the distance or the extinction will alter $p(x,y)$
and so it must be re-normalised.

So far we have not discussed how to deal with errors. To compare with
the data we need the model CMD after it has undergone the same error
processes involved in the observations. It then becomes the
probability density function of obtaining the observed photometry. We
thus consider $\rho(x,y)$, the probability density function of the
pure model CMD, that is, of a star having certain photometry and
$p(x,y)$, the probability density function of the model with errors,
that is, of a star being observed to have certain photometry. The
probability of observing the $i$th star, $p_i = p(x_i,y_i)$ is

\begin{equation}
p_i= \int \rho(x_0,y_0) U(x_i-x_0,y_i-y_0) {\rm d}x_0{\rm d}y_0,
\end{equation}

\noindent
where $ U_(x_i-x_0,y_i-y_0)$ is an error function that varies with
position in the CMD. It has the effect of taking each point in
$\rho(x_0,y_0)$ and spreading it over an error kernel. The natural
choice is a Gaussian:

\begin{equation}
U(x_i-x_0,y_i-y_0)=\frac{1}{2\pi \sigma_x \sigma_y}e^{\frac{-(x_i-x_0)^2}{2\sigma^2_x}}e^{\frac{-(y_i-y_0)^2}{2\sigma^2_y}},
\label{eq:Lmax}
\end{equation}

\noindent
where $\sigma_x(x_0,y_0)$ and $\sigma_y(x_0,y_0)$ are the errors as a
function of position within the CMD. The easiest way to define the
errors is to partition the CMD into a Voronoi Tessellation: each point
adopts the errors of the nearest observation. A better method would be
to analyse the response of the observing apparatus and the reduction
method to artificial stars. We have so far ignored the other principal
source of deviation from an ideal situation: incompleteness at dimmer
magnitudes. If the form of the completeness function is known then it
can be applied to $\rho(x,y)$ before the error-blurring process.

Whether one is interested in maximising the likelihood or applying
Bayesian methods it will be necessary to calculate this likelihood
repeatedly for different star formation rates. It is thus desirable to
minimise the computational load. The error process can be applied to
the pure partial CMDs at the beginning of the analysis. These partial
CMDs must not be normalised individually: the fact that older
populations have fewer stars is important information, it means that
the more massive entities have died. Instead one can record two sets
of data: $p_{ij}$, the value of the $j$th partial CMD at the location
of the $i$th observed star; and $N_j$, the value obtained by
integrating the $j$th CMD over the allowed range of colour and
magnitude. The arrays that define the partial CMDs can then be
ignored and the likelihood becomes

\begin{equation}
\mathcal{L}_1 \propto \prod_{i=1}^{S} \frac{\sum_{j=1}^{M} r_{j} p_{ij}}{\sum_{j=1}^{M} r_{j} N_{j}}.
\end{equation}

\noindent
where  $r_j$ is  the star  formation  rate associated  with the  $j$th
blurred CMD. These must sum to one and the denominator in the fraction
is  required to  normalise each  probability. If the  distance and/or
extinction are varied then the positions of the model stars will shift
and  the $p_{ij}$ will  thus be  different. The  normalisation factors
will also change.

\subsection{The unbinned likelihood: Method 2}

Unfortunately it is not always reasonable to assume a universal error
function. We may have two stars in a similar part of the CMD but with
different photometric errors, a result perhaps of different observing
conditions. The stars may also be so few that we cannot obtain
reasonable error coverage of the CMD. Under such circumstances it
seems natural to calculate the probability of each star given the
model by integrating $\rho(x,y)$ with an error function that
represents the observational uncertainties. This approach goes back to
\citet{1996ApJ...462..672T}, who approximated their model CMD by
drawing from it an equal number of stars to that in the dataset. Their
method had the disadvantage of introducing unnecessary randomness: the
same model CMD will produce different model
populations. \citet{2006MNRAS.373.1251N} improved the method by
instead comparing the data directly with the model CMDs. They defined
the probability of making the $i$th observation to be

\begin{equation}
p_i = \int \rho(x_0,y_0) U_i(x_0-x_i,y_0-y_i)  {\rm d}x_0{\rm d}y_0,
\end{equation}

\noindent
where $U_i(x_0-x_i,y_0-y_1)$ is the error function of the
observation. They defined the statistic $\tau^2$ by analogy with
$\chi^2$ so that

\begin{equation}
\tau^2 = -2 \ln \prod_{i=1}^{S} p_i = -2 \sum \ln p_i ,
\end{equation}

\noindent
and minimised it to find the distances and ages of open clusters. This
approach defines $\mathcal{L}_{2}$, the second definition of the
likelihood.

\begin{equation}
\mathcal{L}_{2}  = \prod_{i=1}^{S} \int \rho(x_0,y_0) U_i(x_0-x_i,y_0-y_i)  {\rm d}x_0{\rm d}y_0.
\end{equation}

If we take the error function to be a Gaussian we have

\begin{equation}
U(x_0-x_i,y_0-y_i)=\frac{1}{2\pi \sigma_{x,i} \sigma_{y,i}}e^{\frac{-(x_0-x_i)^2}{2\sigma^2_{x,i}}}e^{\frac{-(y_0-y_i)^2}{2\sigma^2_{y,i}}},
\end{equation}

where the star is at $x_i$,$y_i$ and has associated errors
$\sigma_{x,i}$ and $\sigma_{y,i}$. This looks very similar to
Equation~\ref{eq:Lmax} but is crucially different: with
$\mathcal{L}_1$, to find the probability of a single observation we
required the errors as a function of position to determine how
different points in $\rho(x,y)$ could be scattered into $p(x,y)$. To
find the probability of a single observation with $\mathcal{L}_2$ we
use its errors alone. The two definitions will be the same if the
errors are everywhere constant and this is true more generally for
invariant error functions that are symmetric about the
observations. In practice they will give similar results if the errors
change over a longer length-scale than the model. If the errors become
very small then the Gaussians become delta-functions. Both definitions
of the probability of an observation then reduce to the value of
$\rho(x,y)$ at the location of the observation.

The appropriate normalisations for $\rho(x,y)$ and
$U_i(x_i-x_0,y_i-y_0)$ require some thought.
\citet{2006MNRAS.373.1251N} normalised the error functions so that
they had the same maximum value, with the aim of recovering
$\tau^2$=$\chi^2$ when the model is a curve and the uncertainties are
only in one dimension. However, an alternative choice, proposed by
\citet{2009MNRAS.399..432N}, is that both the model and the error
functions integrate to one. This is the normalisation that we
adopt. This decision may be put on more rigorous footing by noting
that it is equivalent to treating them as probability distributions
that represent different types of uncertainty. The distribution
$\rho(x,y)$ represents the random uncertainty resulting from a drawing
from the model. By contrast, $U_i(x_i-x_0,y_i-y_0)$ represents the
epistemic uncertainty due to imprecise data. In the terminology of
\citet{Denoeux} it is a Bayesian belief function. The likelihood may
then be considered to be the probability of \emph{fuzzy} data given
the model. As with $p(x,y)$, $\rho(x,y)$ must be normalised over a
fixed area in apparent magnitude and uncorrected colour.

For it to be possible to multiply them together it is necessary that
the belief functions be defined over the same domain as the
model. This may not always be the case. One often has reasons to
impose a cut in apparent magnitude. This may be to obtain a sample
with a high and constant completeness and thus to avoid the
consideration of completeness functions. In addition, dimmer stars not
only often have greater photometric errors than their more massive and
brighter brethren but their long lifetimes make them poor tools for
age determination. One of the outcomes of this cut will be a number of
stars that have have non-trivial parts of their belief functions
outside the region specified by the model. Possible responses are to
either drop these stars from the sample or to perform the integration
over only the area of the CMD allowed by the cuts. We adopt the latter
approach but note that it is not, strictly speaking, correct. One can
imagine the case of a star that, thanks to errors, is above the
magnitude cut, though the area of the model CMD that best matches it is
below. In practice we do not find that this causes problems when we
consider synthetic populations in a later section.

A method for maximising something akin to $\mathcal{L}_2$ in the
context of the inference of star formation histories was proposed by
\citet{2013MNRAS.428..763S}. The principal difference is that they
calculate the probability of observing each star given the model by
\emph{separately} multiplying each isochrone, as considered in a
magnitude-magnitude diagram, with a completeness function and the IMF
and then integrate with the Gaussian error functions. The analogous
approach with our method is to integrate each un-normalised partial CMD
$\rho_j(x,y)$ with each error function $U_i(x_i-x_0,y_i-y_0)$ to get
$q_{ij}$. The probability of an observation is then the normalised
weighted sum of these, with the weights given by the rates
$r_{j}$. The likelihood then becomes

\begin{equation}
\mathcal{L}_2 \propto \prod_{i=1}^{S} \frac{\sum_{j=1}^{M} r_{j} q_{ij}}{\sum_{j=1}^{M} r_{j} N_{j}},
\end{equation}

where, as at the end of Section 2.4, $N_j$ is the value obtained by
integrating the $j$th partial CMD over the allowed range of colour and
magnitude. The advantage of this approach is that one can then
calculate the likelihood for different rates much quicker than if one
had to repeatedly integrate over the CMD.

The second definition of the likelihood, $\mathcal{L}_2$, is always
going to be preferable when one cannot use an overall error function,
although, as we have noted, it may become problematic if cuts are
imposed. We use $\mathcal{L}_2$ in the later sections but briefly
contrast it with $\mathcal{L}_1$ when we consider real data.

\subsection{Overfitting and the choice of temporal bins}

When constructing the partial CMDs we must pick a temporal binning
scheme. The bin boundaries are nearly always chosen to be fixed
advances in the logarithm of time \citep{2002MNRAS.332...91D}. This is to
reflect the decreasing degree by which stellar populations change over
time: massive stars evolve faster and die younger.

We might think that we ought to bin in time as finely as possible so
as to recover all the detail in the star formation history. This is encouraged by the
fact that the maximised likelihood increases monotonically with the
number of bins. This is a more general result: the maximum likelihood
increases with model complexity. Eventually though the model would be
fitting random variations in the data. Overfitting would have
occurred and this is wasteful at best and misleading at worst.

With partial CMDs there is a particular problem in that, as the
time-bin size decreases, adjacent CMDs become more and more
similar. The errors on the rates become increasingly correlated and the
likelihood surface becomes flat. This means that the rates can be
altered considerably without affecting the fitting parameter. The
estimates for the rates are then effectively meaningless. A typical
phenomenon is that, of two adjacent partial CMDs, one is a
slightly better fit to the data. The other rate is then set to zero
and the star formation history takes on an unpleasant jagged form. This phenomenon
may be verified by fitting, for example, the first fifty or so prime
numbers with ${ Ax + Bx^{1.00001}}$ in Mathematica. With no
constraints, A and B have very large values and opposite signs. When
zero values are prevented one is positive and the other zero.

To prevent overfitting we require some sort of application of Occam's
Razor: the simplest (good) explanation is the best. Within the
maximum-likelihood framework we may apply three methods. First,
picking a fitting and a testing set from the data and applying the
methods of bootstrapping or cross-validation. Secondly there are
several information criteria that penalise models with more
parameters. The original of these is the Akaike Information Criterion
\citep{AIC} or AIC.

\begin{equation}
{\rm AIC} = -2 \ln(\mathcal{L}) + 2M.
\end{equation}

\noindent
Instead of maximising the likelihood we minimise the AIC. The solution
is the model that minimises the expected information loss in the
asymptotic limit. It is thus poorly-adapted to small data sets. An
alternative is the Bayesian Information Criterion (BIC) of \citet{BIC},

\begin{equation}
{\rm BIC} = -2\ln(\mathcal{L}) + M\ln(N),
\end{equation}

\noindent
where $N$ is the number of data points, CMD bins in the binned case,
observed stars in the unbinned case, and $M$ is the number of
parameters.

The third method is regularisation, that is, the penalising of models
where the output is not smooth. \citet{2005ESASP.576..171J} considered
the likelihood of a data set given a star formation rate based on very
narrow time bins and maximised it with the aid of a smoothing
parameter $\alpha$. The problem with this approach is that the choice
of smoothing parameter is non-trivial and ad hoc, it is as arbitrary
as the problem of choosing the appropriate number of bins. The errors
on the rates are also then difficult to calculate.

\section{Fitting as Bayesian inference}

We propose a new method. We start by noting that the likelihood is the
probability of the data given the model, whereas the probability of
the model given the data is what is really desired. We thus consider
Bayes' Theorem: instead of just considering the likelihood we combine
it with the prior probability distribution to get the posterior
probability distribution. For parameters $\btheta$, data $D$ and model
$M$, that is, the parametrisation, we have

\begin{equation}
\Pr( {\btheta} \mid D,M) = \frac{\Pr( D \mid {\btheta},M)\Pr({\btheta} \mid M)}{\Pr(D \mid M)}. \label{BAYES}
\end{equation}

\noindent
More generally, for model fitting we have

\begin{equation}
\rm Posterior = \frac{Likelihood \times Prior}{Evidence}.
\end{equation}

\noindent
The aim is to infer the posterior, which is a \emph{probability
  distribution} rather than a point estimate. It represents the belief
of a reasonable person, given both their prior beliefs and the
information provided by the data. This natural interpretation, that
model inference is a problem in probability theory, is the best
justification for the Bayesian approach.

It should be noted that more-or-less Bayesian methods have been
applied to colour-magnitude problems
before. \citet{1996ApJ...462..672T} described their fitting method as
Bayesian but given that it consists of maximising the posterior given
flat priors it is entirely equivalent to maximising the
likelihood. \citet{2005A&A...436..127J} considered open clusters and
calculated the posterior probability of the age of each star given its
photometry.  Their method is also not fully Bayesian: instead of
considering the full posterior distribution they derive confidence
intervals based on analogies with $\chi^2$. Finally,
\citet{2006ApJ...645.1436V} presented a fully Bayesian method for
inferring the posterior distributions of the masses of stars within a
cluster and the cluster parameters (age, distance, metallicity and
extinction). This is a similar but different problem: they assumed a
coeval population and solved for the masses, we assume a common mass
function and solve for the star formation history.

In the remaining part of this section we describe our parametrisation,
consider appropriate priors, and discuss how best to represent the
posterior. We then show how Bayesian methods allow us to avoid
overfitting.

\subsection{The parametrisation}

Star formation in clusters is often observed to have occurred in
bursts \citep{2003MNRAS.341...33K}, the result perhaps of collisions
with gas clouds or other clusters. More generally we expect a star
formation history to exhibit a degree of temporal correlation: if we
know the rate at a particular time then we do not expect it to change
drastically over some small interval. This inspires us to propose a
model whereby the star formation history is represented as a number
$M$ of Gaussians in linear time, each with parameters $\theta_{\rm
  i}$: time before the present day (mean) $t_{i}$, width (standard
deviation) $w_{i}$, and mass $\pi_{i}$.

\begin{equation}
r(t) = \sum_{i=1}^{M} \pi_{i} \mathcal{N}(t;t_{i},w^2_{i}).
\end{equation}

 Our use of the unbinned likelihood means that the model CMD is then
 the probability density function of observing one star. This means
 that the masses assigned to each Gaussian are merely relative weights
 and can be considered to sum to one. The model CMD is produced by
 first generating $B$ partial CMDs by binning the models in linear
 time on a sufficiently small timescale. We used 1 Myr. The linear
 combination is then formed as before with the weights given by the
 function $r(t)$.

\subsection{The prior probabilities}

By default we have no a priori reason to suspect any time is more
likely to have a burst than any other, with the obvious caveat that
the burst time cannot be later than the present day or earlier than
the age of the universe. We thus adopt a uniform prior on $t_{i}$ over
this range.

The widths are scale parameters and to represent our uncertainty
concerning them is more difficult. This becomes apparent as soon as
alternative parametrisations are considered: we could have used the
variance $w_{\rm i}^2$ or the precision $w_{\rm i}^{-1}$ as measures
of the width of a burst. Using a uniform prior with one parameter is
then equivalent to using a non-uniform prior with the
others. Different parametrisations will give different results. We
therefore adopt the Jeffreys prior \citep{Jeffreys46}, which is
constructed so that it has the desirable property of being invariant
under reparametrisation. For a vector of parameters $\btheta$, the
Jeffreys prior is

\begin{equation}
\Pr(\btheta) \propto \sqrt{\mid \mathbfss{I}(\btheta) \mid},
\end{equation}

where $\mid \mathbfss{I}(\btheta) \mid$ is the determinant of the
Fisher information matrix. The Jeffreys prior for the standard
deviation of a Gaussian is $\Pr(w_{\rm i}) \propto 1/w_{\rm i}$.  This
inverse prior represents the fact that we are ignorant as to the
\emph{scale} of the parameter and is equivalent to assuming a uniform
distribution in log time (see Section 3.1 in \citet{Jeffreys} for the
argument that the prior for a semi-infinite parameter should be its
reciprocal). We do however impose a lower limit of 1 Myr on $w_{i}$.

For the masses we use the symmetric Dirichlet distribution with
concentration parameter $\alpha$. It represents the probability of
obtaining a set of masses on the simplex ${\sum \pi_{i} =1}$ when no
component is favoured over another.

\begin{equation}
\Pr(\vec{\pi} \mid M )= \frac{\Gamma(\alpha M)}{\Gamma(\alpha)^M} \prod_{i=1}^M \pi_i^{\alpha - 1}.
\end{equation}

\noindent
The parameter $\alpha$ represents the degree to which we expect the
masses to differ. If $\alpha > 1$ we favour distributions where the
masses are similar whereas if $\alpha < 1$ we suspect that sparse
distributions are more likely. We adopt $\alpha = 1$. We note that
$\alpha =0.5$ is the Jeffreys prior but find that we get almost
identical solutions.

\subsection{The posterior probabilities of the star formation history}

Once we have determined the form of the posterior distribution we then
sample it with Markov Chain Monte Carlo (MCMC) methods. A simple
Metropolis-Hastings \citep{hastings} algorithm is adequate for this
task. The algorithm is initiated with a randomly-chosen set of model
parameters, that is, at a random location in the space defined by the
model parameters. A new, nearby location is then proposed based on a
random jump drawn from a proposal distribution. The ratio of the
posterior probabilities are then compared. If the new location is more
probable than the current location then the new location is adopted
and the corresponding model parameters are recorded. If the new
location is less probable then it is likewise adopted but with a
probability equal to the posterior ratio. If the new location is not
adopted then the model parameters of the current location are recorded
once more.

After initialisation the algorithm moves towards the location of the
bulk of the probability. This process is known as burn-in and the
parameter values in this stage are dependent on the starting-position
and must be ignored. Once the algorithm has reached the location of
the bulk of the probability it wanders about this subset of the
parameter space. In doing so it recovers an increasingly-accurate
estimate of the posterior probability distribution. The algorithm may
be said to have converged when the output posterior is the same, to a
chosen degree of precision, for different starting locations.

The evaluation of the posterior probability at a particular set of
parameters requires, for the blurred likelihood $\mathcal{L}_1$, the $SB$
relative probabilities that correspond to the value of the $B$ blurred
partial CMDs at the location of the $S$ stars, and the function
$r(t)$. The integrated likelihood $\mathcal{L}_2$, which is the one we
principally use, requires the $SB$ relative probabilities
corresponding to the integration of the $S$ stellar error functions
with the $B$ pure partial CMDs, and the function $r(t)$. Only $r(t)$
must be recalculated each time for either method and thus the
algorithm can accumulate enough samples for good convergence very
quickly. This is not altered by the addition of more filters. The
principal effect of fitting a higher-dimensional structure that
involves more than two filters is that generating the models becomes a
lengthier process.

The posterior distribution of the star formation history has $3M$
dimensions and is difficult to represent. One approach is to take
several random drawings from the distribution. This gives some idea of
the probable nature of the star formation history and its variability
but does not incorporate all the information provided by the
posterior. The approach we take is to determine the posterior
distribution of the rate $r(t)$ by recording the rate at each point on
the Markov Chain. The mean rate at a particular time is then easily
calculated. To find the uncertainties we reflect that, in Bayesian
terms, they define a credible interval in which we believe that the
rate has a certain probability of being. The more compact the interval
the more informative it is. We thus adopt the mode as our best
estimate of the rate and then construct the lower and upper limits
about it such that they form the most compact interval in which the
rate has a 68 per cent chance of being. This is an arbitrary number
but allows an intuitive connection with the idea of $1\sigma$ error
bars. If the posterior distributions of the rates are reasonably
normal in appearance then this approach offers an adequate
representation.

\subsection{The posterior probabilities of the parameters}

 It is important to note that the labelling of the Gaussians is not in
 itself intrinsically important to the inferred star formation history. The likelihood is
 invariant under label-switching. This causes the posterior
 distribution to have $M!$ identical modes. The probability density
 function for each parameter can be found by summing the full
 posterior distribution over the other parameters. This process is
 known as marginalisation and assuming full convergence, each
 parameter of the same type has the same marginal distribution.

The Gaussian burst parametrisation may be no more than a convenient
way of generating a temporally-correlated star formation history. If
however we believe that we have identified individual bursts then it
may be of interest to know their parameters. There are several methods
by which this may be attempted \citep{jasra}. The easiest way is to
process the MCMC output to reorder the bursts by time to impose an
artificial identifiability constraint. The marginal probabilities for
a certain burst's parameters are then those of the burst that has a
certain relative position given the presence of the others. We find
that this generally offers only a limited solution and that the
posterior marginal distributions are still multimodal. It is easy, for
example, to conceive of a scenario where one pair of bursts might be
best distinguished by their times and another, with similar times, by
their widths or masses.

Another method, proposed by \citet{marin}, is to make use of the set of
parameters in the MCMC output that gave the highest posterior
probability. This is the maximum a posteriori (MAP) set of parameters
and is the Bayesian analogue of the maximum-likelihood solution. It is
usually an extreme case and may be unrepresentative of the location of
the bulk of the probability. For this reason we generally do not use
it when considering the overall star formation history. Nevertheless
it provides a standard by which the other samples can be
compared. \citet{marin} suggest that one should relabel the MCMC
output so that each sample resembles the MAP estimate as much as
possible according to some chosen distance metric. We do not find that
this method removes the symmetry from the posterior distributions any
better than an identifiability constraint.

We instead adopt the iterative relabeling algorithm approach of
\citet{stephens1}. In this, one first identifies a loss function based
on the MCMC output. One then relabels each sample to minimise the loss
function and repeats until convergence. We start from the premise
that, if the number of bursts has good posterior support, there should
exist a permutation of the MCMC outputs such that the posterior
distribution of each parameter is reasonably close to a normal
distribution. We first calculate the mean $\mu_{\rm i}$ and variance
$\sigma_{\rm i}^2$ of the parameters given the current set of
orderings. We then permute each MCMC sample such that the product of
the values of the normal distribution with these
means and variances and at each parameter value is minimised. This is the loss function

\begin{equation}
\mathcal{L}_{0}= -\log (\prod^{3M}_{i=1} \mathcal{N} (\theta_i; \mu_i, \sigma_i^2)),
\end{equation}

\noindent
where the product is over the $3M$ model parameters: the times, widths
and masses of the $M$ bursts. We recalculate the new means and
variances and repeat. This algorithm converges on the relabelling
scheme that yields the most normal-like form of the posterior
distributions given the starting point. A global optimum can be
ensured by running the algorithm repeatedly from a number of different
starting points and taking the output that gives the most normal-like
behaviour overall.

We find that this approach usually removes the symmetry from the
posterior distributions and that any multimodality can then be
considered as evidence for an alternative model. We note that the full
marginal distributions are not conveniently presented and that it is
helpful to reduce them to point estimates and uncertainties. Modes and
credible intervals usually suffice.

\subsection{The classification probabilities}

Another consequence of a belief that the bursts are physically
meaningful is a desire to ascertain which stars belong to which
burst. This can be accomplished by reporting $\mathbfss{U}$, the matrix of
the classification probabilities of star $i$ being due to burst
$j$. The matrix $\mathbfss{U}$ may be estimated from the MCMC output by considering
$\mathbfss{V}(\btheta)$, the classification probabilities given
a set of parameters $\btheta$ from an MCMC sample. The
unbinned likelihood of obtaining star $i$ from burst $j$ is
$f_{i}(\btheta_{j})$.

\begin{equation}
V_{ij}(\btheta) = \frac{f_{i}(\btheta_{j})}{\sum_{l=1}^{M} f_{i}(\btheta_{l})}.
\end{equation}

\noindent
The matrix $\mathbfss{U}$ is then given by

\begin{equation}
U_{ij} = \frac{1}{R} \sum_{s=1}^S V_{ij}(\btheta_{s}),
\end{equation}

\noindent
where there are $R$ MCMC samples. We calculate the classification
probabilities for each star and assign them to the most probable burst
if this probability is greater than 0.75. Otherwise their origin is
considered to be unclear and they are labelled accordingly.

\subsection{Model comparison}

Instead of the number of temporal bins, the choice of model now
becomes the number of bursts. We are unlikely to know a priori the
appropriate number of bursts and so the best set of parameters may be
said to be `unknown unknowns': we don't know their values and don't
know that we ought to. This Rumsfeldian\footnote[1]{``There are known
  knowns; there are things we know we know.  We also know there are
  known unknowns; that is to say, we know there are some things we do
  not know.  But there are also unknown unknowns - the ones we don't
  know we don't know.''  \\ Donald Rumsfeld, US Secretary of Defense,
  12th February 2002.}  problem may be addressed by considering the
joint posterior probability of both the model index ($M$, the number
of bursts) and the parameters ${\btheta}$,

\begin{equation}
\Pr( M, \btheta \mid D) \propto \Pr(M) \Pr(\btheta \mid M) \Pr( D \mid M, \btheta)  \label{BAYES2}.
\end{equation}

\noindent
${\rm \Pr({\btheta} \mid M)}$ is the prior probability of the
parameters and ${\rm \Pr( D \mid M, {\btheta})}$ is the
likelihood. The model prior ${\Pr(M)}$ represents our opinion as to
how probable different numbers of bursts are. If we think that the
number of bursts could be anywhere between one and infinity then there
is an argument for using ${\Pr(M)} \propto 1/M$ in the same spirit as
the Jeffreys prior on the widths. This level of prior ignorance is not
really applicable here. The aim of this process is primarily to find
the minimum number of bursts that will represent the data properly:
any more would be effectively superfluous and would merely model
random error. Each burst represents three additional parameters and,
for populations of hundreds to thousands of stars, it would be
unreasonable to use more than a few dozen parameters for modelling
purposes.

We therefore follow \citet{greenrichardson} in adopting a uniform
prior between $M=1$ and $M=M_{\rm max}$, with $M_{\rm max}$ chosen
with the assumption that the posterior probability associated with it
will be low. If this is not the case then the analysis is
repeated with a higher value.

The joint posterior may be factorised as

\begin{equation}
\rm \Pr( M, {\btheta} \mid D) \propto \Pr(M \mid D) \Pr({\btheta} \mid M, D)  \label{BAYES3}.
\end{equation}

\noindent
The first probability, ${\rm \Pr(M \mid D)}$, is the posterior probability
of the model. The second is the more familiar posterior probability of
the parameters given the model. This means that, given the joint
posterior, we can marginalise over the parameter values to find the
model posterior. We can then pick the model with the greatest
posterior probability and use its probability distribution for the
parameters.

The Reversible Jump MCMC method of \citet{green} offers a way to
recover the joint posterior. The essence of the method is to consider
the Metropolis-Hastings proposals to be on a space that includes model
variability and to adjust the acceptance ratios accordingly. We thus
use the following simple method based on the work of
\citet{stephens2}. The Markov chain is able to perform three moves: an
ordinary shift move within the current model, a birth move that adds a
new Gaussian, its parameters drawn from the prior, and a death move,
in which a random Gaussian is removed. Because the masses are obliged
to sum to one, a transdimensional move will affect them all; the other
parameters are not affected. We have, for a birth move that adds a new
burst with a mass of ${\pi^{(M+1)}_{(M+1)}}$,

\begin{equation}
\pi^{( M+1)}_{(i)}= \pi^{(M)}_{(i)}(1-\pi^{(M+1)}_{(M+1)}),
\end{equation}

\noindent
where the superscript indicates the total number of bursts and the
subscript the burst number. A death move similarly rescales the masses
of the surviving bursts. For ordinary Metropolis-Hasting moves within
a fixed model the acceptance ratio a is

\begin{equation}
\rm a(x,y)=\min \left( 1,\frac{P(x)Q(x \mid y)}{P(y)Q(y \mid x)} \right),
\end{equation}

\noindent
where $x$ and $y$ represent two locations in the parameter space, P
the posterior probability and Q the proposal distribution. In the
transdimensional setting and for the case of moving between a model
with one parameter to one with two this becomes

\begin{equation}
a_{12}=\min \left( 1,\frac{P(2,(\theta_1,\theta_2))c_{21}}{P(1,\theta) q(u) c_{21}} \left| \frac{\partial(\theta_1,\theta_2)}{\partial(\theta,u)} \right| \right).
\end{equation}

The variables ${c_{12}}$ and ${c_{21}}$ are the probability of
proposing the forward and reverse moves respectively. Note that if we
move from $M$ to $M+1$ bursts, the probability of moving back to the
original state is the death rate multiplied by $1/(M+1)$. The
one-parameter state had value ${\theta}$, to which we apply the random
variable $u$ to get the new state $\theta_1,\theta_2$. The prior
probability of $u$ is $q(u)$. The final element is the Jacobian of the
transformation. With our model the masses of the bursts are the only
original variable that is changed and this only depends on the new
mass. The Jacobian thus becomes ${(1-\pi^{(M+1)}_{(M+1)})^M}$ and the
acceptance ratio for a birth is

\begin{equation}
a_{\rm birth}=\min \left( 1,\frac{P_{\rm new} (d/(M+1)}{P_{\rm old} Q b}  (1-\pi^{(M+1)}_{(M+1)})^M \right),
\end{equation}

\noindent
where Q is the joint prior on the parameters of the new burst, $P_{\rm
  old}$ is the posterior probability of the old state with $M$ bursts
and $P_{\rm new}$ that of the new state with $M+1$ bursts. The
quantities $b$ and $d$ are the probabilities of the algorithm
proposing a birth or death move. The acceptance ratio for a death move
is the inverse of this expression, with the difference that the prior
$Q$ is that of the burst that is to be removed.

With this transdimensional Metropolis-Hastings method we can recover
the joint model posterior over the parameters and the model
index. More complicated models are automatically penalised by this
process because the prior becomes spread over more dimensions. The
prior probability of a given set of parameters is then reduced, even
though the likelihood will be greater. The algorithm thus applies
Occam's Razor and ends up with the minimum number of bursts required
to explain the data.

We can also make direct comparison between models of different types:
all that is required is a jump move and an associated Jacobian. It was
suggested to us that exponentials might be a better way of
representing the star formation history than Gaussians. This is not
unreasonable: there is no particular reason to expect a burst to be
symmetrical.

We tested this by adding a fourth type of Metropolis-Hastings move,
model switching between Gaussians and exponentials. We identified the
mass, mean time and width of the Gaussians with the mass, start time
and time constant of the exponentials. This transformation has a
Jacobian of 1 and, as we used the same Jeffrey's priors, the
acceptance probabilities are just the ratio of the posterior
probabilities. We found that, for the observed data that we
considered, the Gaussians received about three times more posterior
support than the exponentials. This justifies their use. We note in
passing that jumps between different models can suffer from poor
acceptance rates, particularly if the models are very different. If
this is the case then we recommend the use of Diffusive Nested
Sampling \citep{Brewer}.

A final point is that, as \citet{greenrichardson} observed,
model switching greatly assists the mixing of the Markov chain within
a particular model. The random addition and subtraction of Gaussians
means the chain is able to traverse the posterior landscape in jumps
and is much less likely to get stuck around a single mode.

\section{The stellar models}

Our synthetic partial CMDs were produced with output from the
Cambridge ${\sc \rm STARS}$ code. This was originally developed by
Peter Eggleton in the 1960s \citep{Eggleton}. It uses a
non--Lagrangian mesh, where the mesh function ensures that the points
are distributed so that quantities of physical interest do not vary by
a large amount in the intervals. The code has been gradually improved
and updated and the work in this paper is based on the code described
by \citet{Stancliffe} and references therein.

Convection is included in the code via the mixing length theory of
\citet{BohmVitense} with a solar-calibrated mixing length parameter
of $\alpha=2.0$.  We use the mass-loss scheme described by
\citet{2008MNRAS.384.1109E}. For main-sequence OB stars the
mass-loss rates are calculated according to
\citet{2001A&A...369..574V} and for all other non-Wolf-Rayet stars we
use the rates of \citet{1988A&AS...72..259D}, where the metallicity
fraction by mass ($Z$) scaling goes as ${(Z/Z_{\odot})^{0.5}}$. This
theory is older but the the rates have been recently tested for red
supergiants and been shown to be still the least-worst available
\citep{2011A&A...526A.156M}. We use the rates of
\citet{2000A&A...360..227N} for Wolf-Rayet stars but scale them
according to metallicity again by ${(Z/Z_{\odot})^{0.5}}$. The value of
the Wolf-Rayet scaling exponent is somewhat uncertain but some degree
of scaling is required to give better agreement with observations
\citep{2006A&A...452..295E}.

We created a library of evolution models with different values of
$Z$. The fractions of hydrogen and helium were determined on the
assumption of constant helium enrichment from the primordial condition
of $X$=0.75, $Y$=0.25. We then calibrated to a Solar composition of
$X$=0.70, $Y$=0.28 and $Z$=0.02 so that $Y=0.25+1.5Z$. The masses
start at ${0.5\,\rm M_{\odot}}$ and increase every ${0.1\,\rm
  M_{\odot}}$ to ${3\,\rm M_{\odot}}$, every ${0.2\,\rm M_{\odot}}$ to
${12\,\rm M_{\odot}}$, every ${0.5\,\rm M_{\odot}}$ to ${20\,\rm
  M_{\odot}}$, every ${1\,\rm M_{\odot}}$ to ${50\,\rm M_{\odot}}$ and
every ${2\,\rm M_{\odot}}$ to ${100\,\rm M_{\odot}}$.

These models were processed to generate observed magnitudes with the
BaSeL V3.1 model atmosphere grid \citep{Westera} and the relevant
broad-band filter functions. The magnitudes were then interpolated
for the values at different fractions through the life of the stars
and these were then in turn interpolated between the masses. This
meant that interpolation in mass was between models at equivalent
stages in their lives.The relevant parts of each interpolated stellar
evolution track were then assigned to the appropriate time bin with a
weighting set by the model timestep and the IMF.

\section{Application to synthetic populations}

We began by testing our method with synthetic populations with a known
star formation history. We used Z=0.008 for the metallicity,
principally because we use the same models in the following section
when we consider real data. These observations were in F435W, F555W
and F814W, which for brevity we refer to as B, V and I, although at no
point did we actually convert to the Johnson-Cousins filters. We ran
the Markov Chain until six million samples were obtained and
discarded the first 50,000 of these to avoid the burn-in phase. The
posterior distributions were tested for consistency by running the
program repeatedly from different starting points. The probability of
a birth move was set to 0.2, a death move to 0.2 and a shift move to
0.6.

\subsection{Population 1: Two Gaussian bursts}

\begin{figure}
\centering
\includegraphics[width=0.5\textwidth]{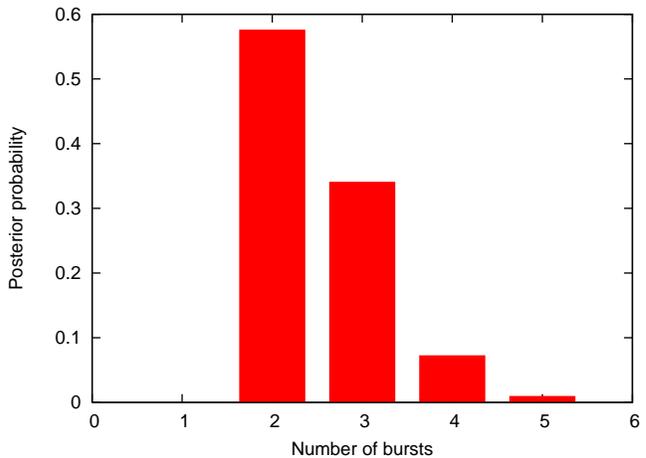}
\caption{The posterior probabilities of the number of bursts used to fit Population 1. Two bursts give the best fit.}
\label{fig:m1}
\end{figure}

\begin{figure} 
\centering
\includegraphics[width=0.5\textwidth]{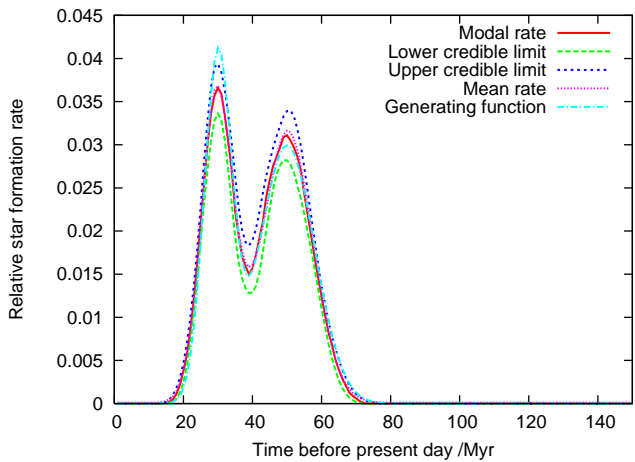}
\caption{The inferred star formation history of Population 1. The posterior probability of the star formation rate at each point in time is summarised by its mean, mode and minimum 68 per cent credible interval. The generating function is the star formation history from which the test population was produced.}
\label{fig:r1}
\end{figure}

\begin{figure}
\centering
\includegraphics[width=0.5\textwidth]{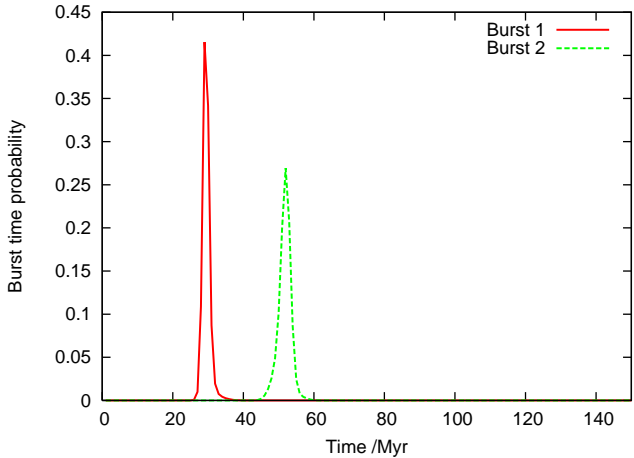}
\caption{The posterior probabilities of the times of Population 1.}
\label{fig:t1}
\end{figure}

\begin{figure}
\centering
\includegraphics[width=0.5\textwidth]{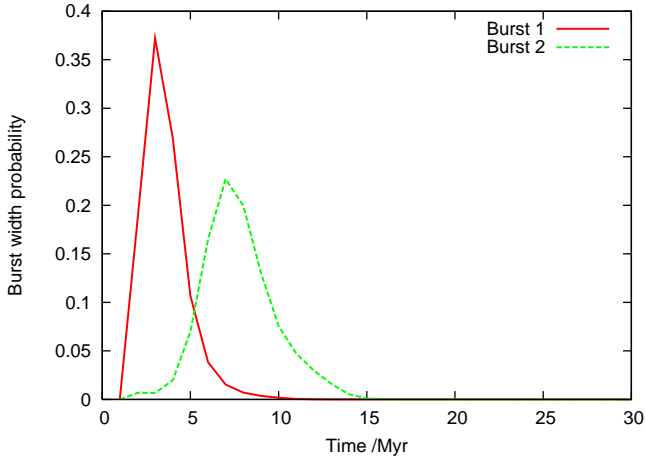}
\caption{The posterior probabilities of the widths of Population 1.}
\label{fig:w1}
\end{figure}

\begin{figure} 
\centering
\includegraphics[width=0.5\textwidth]{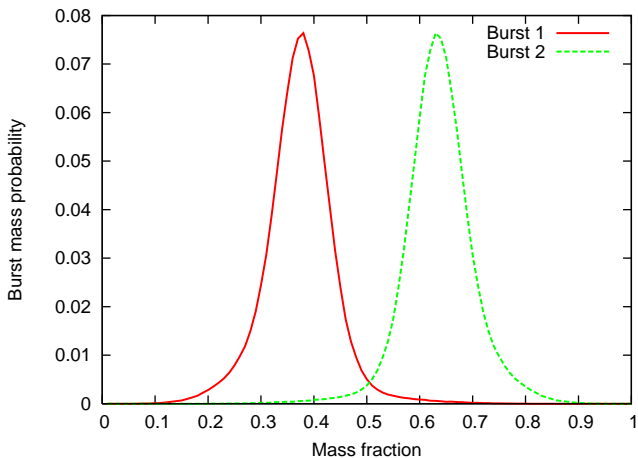}
\caption{The posterior probabilities of the mass fractions of Population 1.}
\label{fig:mm1}
\end{figure}

\begin{figure} 
\centering
\includegraphics[width=0.5\textwidth]{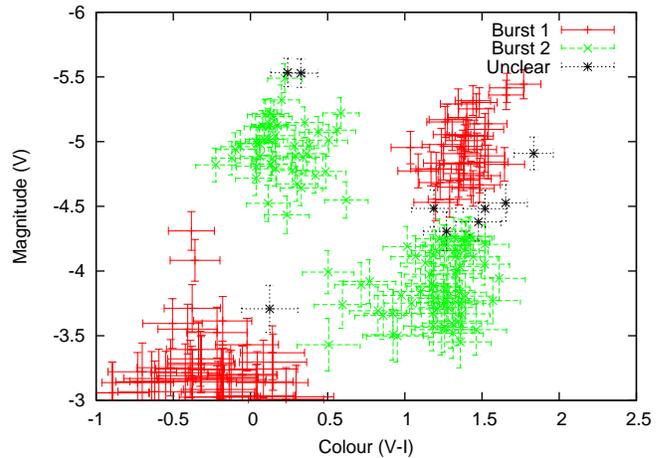}
\caption{The CMD with classified stars of Population 1.}
\label{fig:c1}
\end{figure}

The least demanding test is to see whether the program can fit a
stellar population that is genuinely composed of Gaussian bursts. We
considered a simple star formation history consisting of two bursts
and the parameters we chose are listed in Table~\ref{tab:cluster1}. We
formed the synthetic CMD in $V$ from $-3$ to $-8$ and in $V-I$ from
$-1$ to $-3$ and applied a simple but somewhat arbitrary error
function that depended inversely on magnitude. The error in the
magnitude is $0.5/(-2-M_{\rm V})$ and the error in the colour is the
same for a given star. We also assumed full completeness for the sake of
simplicity and this approach is used in the three later synthetic
populations as well. We then created a population of 200 stars from
this probability density function by applying the rejection
method. The approximate limit of detection set by the stellar models
is 300 Myr; stars older than this are either dead or below the
magnitude cut-off

\begin{table}
\caption{The input parameters of Population 1.}
\begin{tabular}{| l | l | l | l |}
  \hline
   Burst & Time /Myr  & Width  /Myr & Mass fraction  \\
  \hline
  1 & 30 & 4 & 0.4 \\
  2 & 50 & 8 & 0.6 \\
  \hline
\end{tabular}
\label{tab:cluster1}
\end{table}

We ran the MCMC code and found that, as expected, the star formation
history is best fitted with two bursts (Fig.~\ref{fig:m1}). The modal
values of the probability density function of the rate
(Fig.~\ref{fig:r1}) are a reasonable match for the generating
function. The mean rate is very similar to the mode and the
probability density functions for the rates are approximately
normal. The probability density functions for the times, widths and
masses are shown in Fig.~\ref{fig:t1}, Fig.~\ref{fig:w1} and
Fig.~\ref{fig:mm1} respectively. Finally we present the CMD with the
stars labelled according to their classification probabilities
(Fig.~\ref{fig:c1}). The first burst has produced the main sequence
turn-off and the brighter red supergiants, the second has produced the
dimmer red supergiants and the helium-burning stars on the blue loop.

If we consider the probability density functions of the parameters we
see that they are unimodal and reasonably normal in appearance. We
adopt the mode as our best estimate and present them together with the
minimal 68 per cent credible intervals in
Table~\ref{tab:cluster11}. Happily they are in agreement with the
input parameters. It is notable that the percentage error in the times
is less than that of the masses and widths. To some extent there is a
degeneracy in the latter two parameters: the star formation rate at
the centre of the Gaussian can be reduced by either increasing the
width or reducing the mass. We note that, unsurprisingly, if the number
of stars is increased the uncertainties and the differences between the
estimates and the input parameters are reduced. With $10^4$ stars
these become less than 1 Myr for the times and widths and 0.01 for the
mass fractions.

\begin{table}
\caption{The estimated parameters of Population 1.}
\begin{tabular}{| l | l | l | l |}
  \hline
   Burst & Time /Myr  & Width  /Myr & Mass fraction  \\
  \hline
  1 & $28^{+2}_{-1}$ & $2^{+2}_{-1}$ & $0.37^{+0.05}_{-0.05}$ \\
  2 & $51^{+1}_{-2}$ & $6^{+3}_{-1}$ & $0.62^{+0.05}_{-0.05}$\\
  \hline
\end{tabular}
\label{tab:cluster11}
\end{table}

\begin{figure} 
\centering
\includegraphics[width=0.5\textwidth]{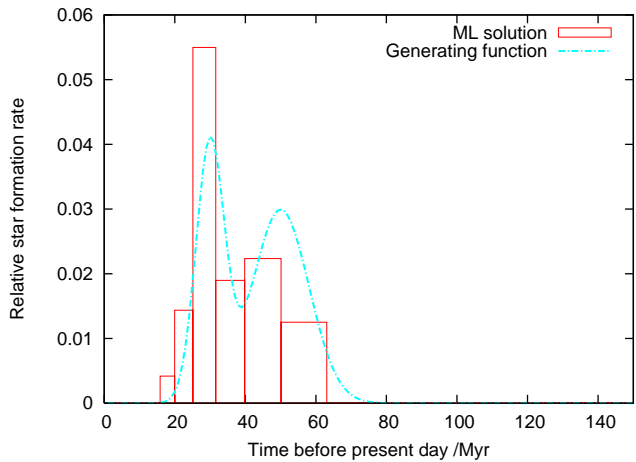}
\caption{The histogram-based maximum likelihood ($\mathcal{L}_2$) star formation history of Population 1.}
\label{fig:r1ml}
\end{figure}

By way of contrast we also provide the star formation history obtained
by maximising $\mathcal{L}_2$ with partial model CMDs that were binned
according to advances of 0.1 in the base 10 logarithm of time. This
histogram-based method is the same as that of
\citet{2002MNRAS.332...91D}; the only difference is that we maximise
the unbinned integrated likelihood $\mathcal{L}_2$, the same
likelihood used in our Bayesian method, instead of the binned Poisson
likelihood $\chi^2_\lambda$. To effect the maximisation we used a
simple simulated annealing algorithm and the result is shown in
Fig.~\ref{fig:r1ml}. The match is reasonable but the shortcomings of
the method are apparent: it is unable to fit a continuous function or
to automatically identify the existence of two different components.

%Finally it may be of interest to see how the uncertainties are reduced
%as the number of stars is increased. We repeated the MCMC analysis
%with the same star formation history but with $10^2$, $10^3$, $10^4$
%and $10^5$ stars. The error in the measured parameters was less
% The results are shown in Table~\ref{tab:cluster111}. It
%is worth making the point that the analysis became very slow when the
%number of stars exceeded ten thousand. With so much data the loss of
%information induced by binning the CMD is small and the gain in speed
%can justify it. The unbinned likelihood is best applied to smaller
%populations.
%
%\begin{table}
%\caption{The estimated parameters for different numbers of stars.}
%\begin{tabular}{| l | l | l | l |}
%\hline
%Burst & Time /Myr  & Width  /Myr & Mass fraction  \\
%\hline
%100 stars \\
% 1 & $28^{+2}_{-1}$ & $2^{+2}_{-1}$ & $0.37^{+0.05}_{-0.05}$ \\
% 2 & $51^{+1}_{-2}$ & $6^{+3}_{-1}$ & $0.62^{+0.05}_{-0.05}$\\
%1,000 stars \\
% 1 & $28^{+2}_{-1}$ & $2^{+2}_{-1}$ & $0.37^{+0.05}_{-0.05}$ \\
% 2 & $51^{+1}_{-2}$ & $6^{+3}_{-1}$ & $0.62^{+0.05}_{-0.05}$\\
%10,000 stars \\
% 1 & $28^{+2}_{-1}$ & $2^{+2}_{-1}$ & $0.37^{+0.05}_{-0.05}$ \\
% 2 & $51^{+1}_{-2}$ & $6^{+3}_{-1}$ & $0.62^{+0.05}_{-0.05}$\\
%100,00 stars \\
% 1 & $28^{+2}_{-1}$ & $2^{+2}_{-1}$ & $0.37^{+0.05}_{-0.05}$ \\
% 2 & $51^{+1}_{-2}$ & $6^{+3}_{-1}$ & $0.62^{+0.05}_{-0.05}$\\
% \hline
%\end{tabular}
%\label{tab:cluster11}
%\end{table}

\subsection{Population 2: Three delta-functions}

\begin{figure}
\centering
\includegraphics[width=0.5\textwidth]{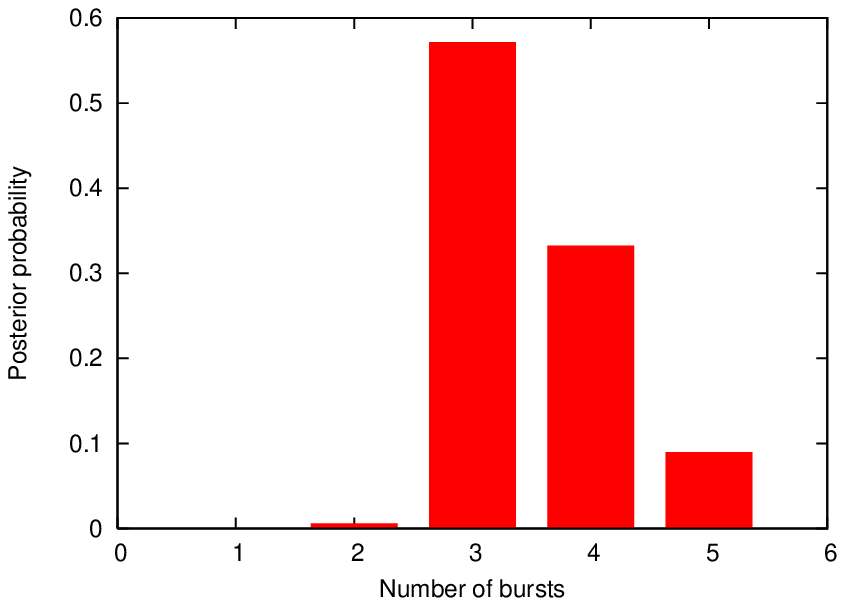}
\caption{The posterior probabilities of the number of bursts used to fit Population 2. Three bursts give the best fit.}
\label{fig:mx}
\includegraphics[width=0.5\textwidth]{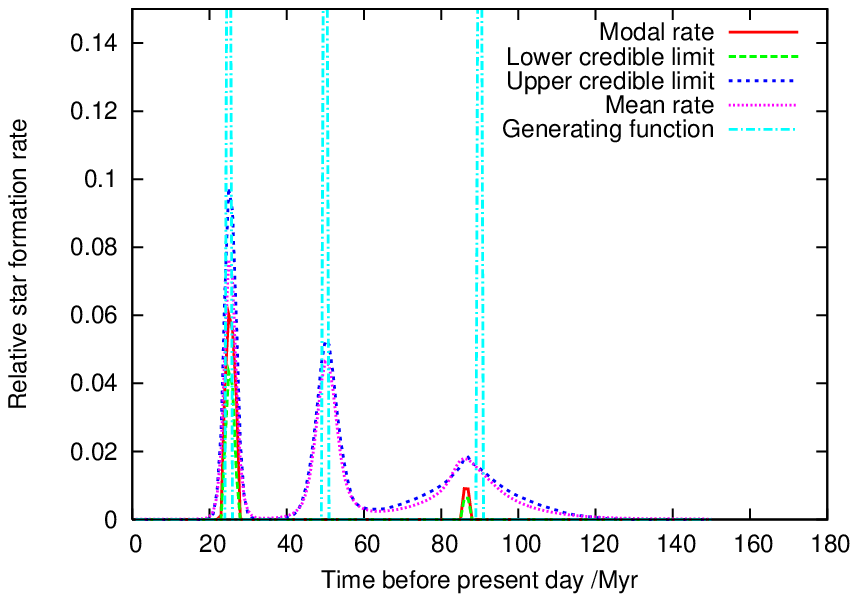}
\caption{The inferred star formation history of Population 2. The posterior probability of the star formation rate at each point in time is summarised by its mean, mode and minimum 68 per cent credible interval.}
\label{fig:rx}
\includegraphics[width=0.5\textwidth]{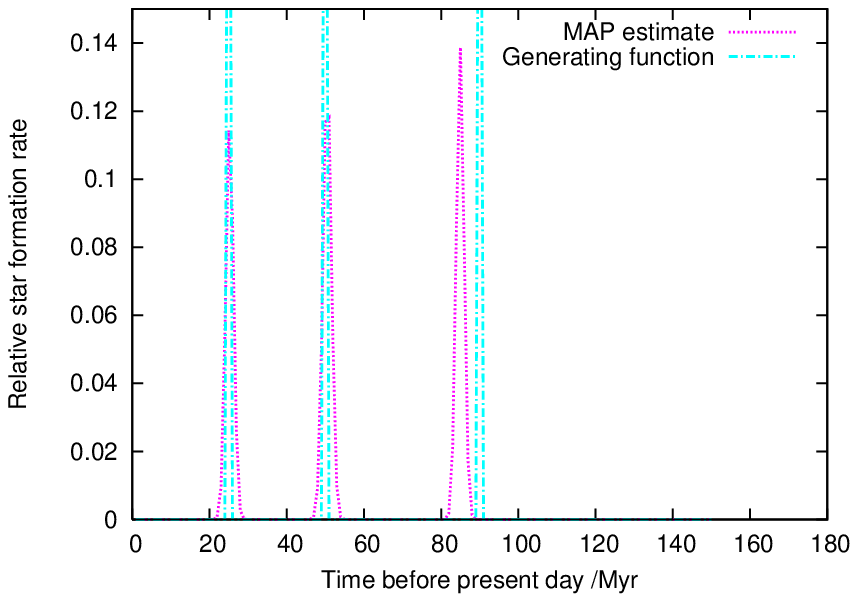}
\caption{The maximum a posteriori (MAP) star formation history of Population 2.}
\label{fig:rxx}
\end{figure}

Three delta-function bursts provided our second test case. We used a
star formation history consisting of delta-functions at 25, 50 and 90
Myr and generated 150 stars. We found, as expected, that three bursts
provide the best fit (Fig.~\ref{fig:mx}). The inferred parameters are
given in Table~\ref{tab:cluster33} and the rate is shown in
Fig.~\ref{fig:rx}.

The modal rate is zero for much of the plot. This is a consequence of
the fact that the MCMC samples consist of very narrow bursts. The
posterior distribution of the rate at a given time thus has most of
the probability at zero, with the rest at very high values
corresponding to the occasional presence of a narrow burst. This
highly-non normal distribution is difficult to represent adequately
with the summary statistics plot. We therefore also present the
maximum a posteriori (MAP) solution, which in this case is a much
better indication of the nature of the generating function
(Fig.~\ref{fig:rxx}). The bursts all have the minimum possible width
of 1 Myr and would have been narrower still if the program had allowed
it. In general the MAP estimate is an extreme and unrepresentative
case but with this example the generating function is extreme as
well. Random drawings from the posterior distribution are in most
cases similar to this but with some scatter in the times and the masses.

 If we consider the burst parameters we find that, within the
 uncertainties, the times are correctly identified and the masses are
 equal. The uncertainties in the times and the widths are highest for
 the oldest burst. This is a consequence of the fact that stellar
 populations change more slowly with increased time and hence their
 resolving power is weaker. The older stars are also dimmer and thus
 have bigger photometric errors.

\begin{table}
\caption{The estimated parameters of Population 2.}
\begin{tabular}{| l | l | l | l |}
  \hline
   Burst & Time /Myr  & Width  /Myr & Mass fraction \\
  \hline
  1 & $25^{+1}_{-1}$ & $1^{+1}_{-1}$ & $0.28^{+0.05}_{-0.03}$ \\
  2 & $50^{+1}_{-2}$ & $1^{+1}_{-1}$ & $0.33^{+0.05}_{-0.06}$\\
  3 & $85^{+5}_{-2}$ & $1^{+12}_{-1}$ & $0.36^{+0.06}_{-0.05}$\\
  \hline
\end{tabular}
\label{tab:cluster33}
\end{table}

\begin{figure} 
\centering
\includegraphics[width=0.5\textwidth]{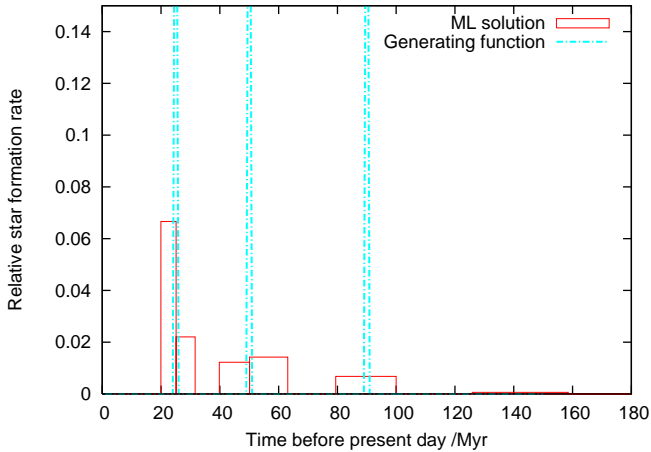}
\caption{The histogram-based maximum likelihood ($\mathcal{L}_2$) star formation history of Population 2.}
\label{fig:r2ml}
\end{figure}

As before we also provide the histogram solution
(Fig.~\ref{fig:r2ml}). The three bursts are identified, the times are
correct and the relative allocation of star formation appears
approximately equal. However, the increasing size of the temporal bins
means that the rate is depicted as declining. The delta-function
nature of the bursts is impossible to ascertain.

\subsection{Population 3: A more complicated history}

Hitherto we considered populations where the individual Gaussians had
some physical meaning, that is, they corresponded to different
bursts/events. We now present a star formation history where the
Gaussians are only considered to be a useful parametrisation. Choosing
the number of Gaussians is then not so much a matter of identifying
different populations as selecting the appropriate number of
parameters with which to fit the star formation history.

\begin{figure}
\centering
\includegraphics[width=0.5\textwidth]{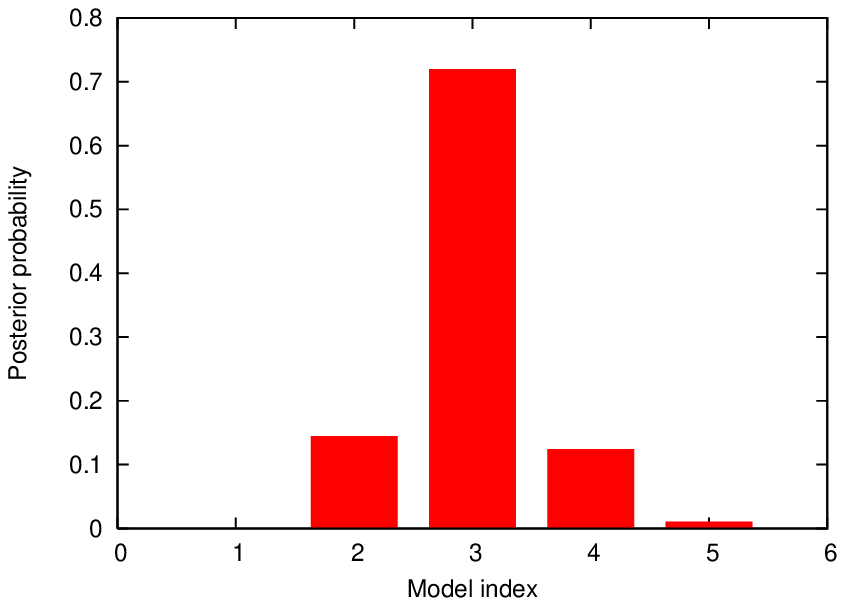}
\caption{The posterior probabilities of the number of bursts used to fit Population 3. Three bursts give the best fit.}
\label{fig:my}
\includegraphics[width=0.5\textwidth]{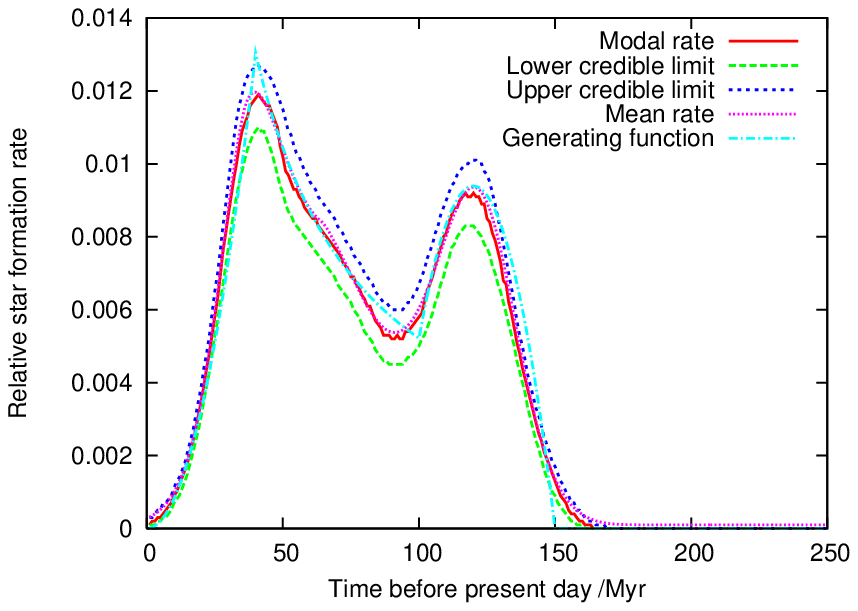}
\caption{The inferred star formation history of Population 3. The posterior probability of the star formation rate at each point in time is summarised by its mean, mode and minimum 68 per cent credible interval}
\label{fig:ry}
\includegraphics[width=0.5\textwidth]{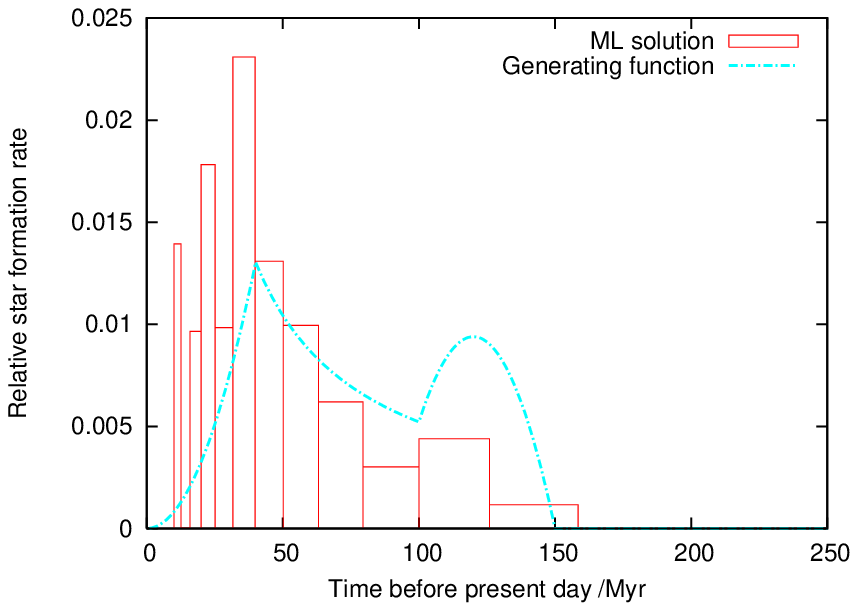}
\caption{The histogram-based maximum likelihood ($\mathcal{L}_2$) star formation history of Population 3.}
\label{fig:r3ml}
\end{figure}

We chose a piecewise continuous functional form $r(t)$ as the star
formation history,

\begin{equation}
r(t) \propto t^2,\,\, 0<t<40,
\end{equation}
\begin{equation}
\propto t^{-1},\,\, 40<t<100,
\end{equation}
\begin{equation}
\propto -At^2+ Bt - C,\,\,100<t<150,
\end{equation}

\noindent
where $t$ has units of Myr, $A=32/25$, $B=1536/5$ and $C=17280$. We
generated 1000 stars from the CMD and applied the MCMC program. We
found that three bursts are required for the best fit
(Fig.~\ref{fig:my}) and that the inferred rate is a reasonable match
to the generating function (Fig.~\ref{fig:ry}). The only difficulty lies
in the proper reproduction of the sharp peak at 40 Myr.

The histogram fit is shown in Fig.~\ref{fig:r3ml} and makes a poor
contrast. The overall form is correct but the small bins at recent
times have high uncertainties. A better match would probably be
obtained by using fewer time bins.

\subsection{Population 4: Direct model comparison with a flat star formation history}

We finally considered a flat star formation rate out to $80\, \rm Myr$. The
Gaussians are very poorly-adapted to fitting such a history and a
rectangular function would be far more appropriate. We demonstrated
the power of the Reversible Jump approach by adding a fourth MCMC move,
jumping between Gaussians and rectangular functions. For the latter,
the time of a rectangle is that of its centre and the equivalent of
the Gaussian standard deviation is the half-width.

\begin{figure}
\includegraphics[width=0.5\textwidth]{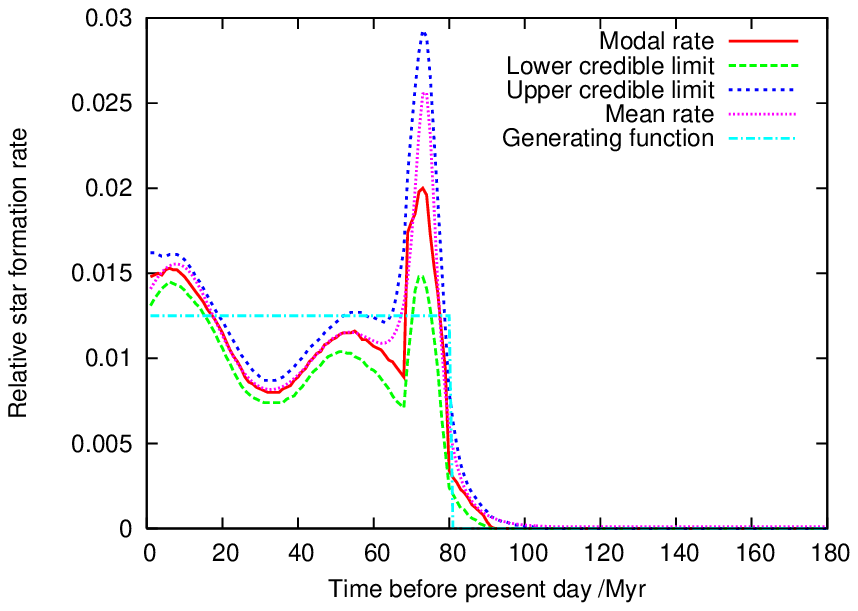}
\caption{The inferred star formation history of Population 4 with Gaussians.}
\label{fig:r2}
\includegraphics[width=0.5\textwidth]{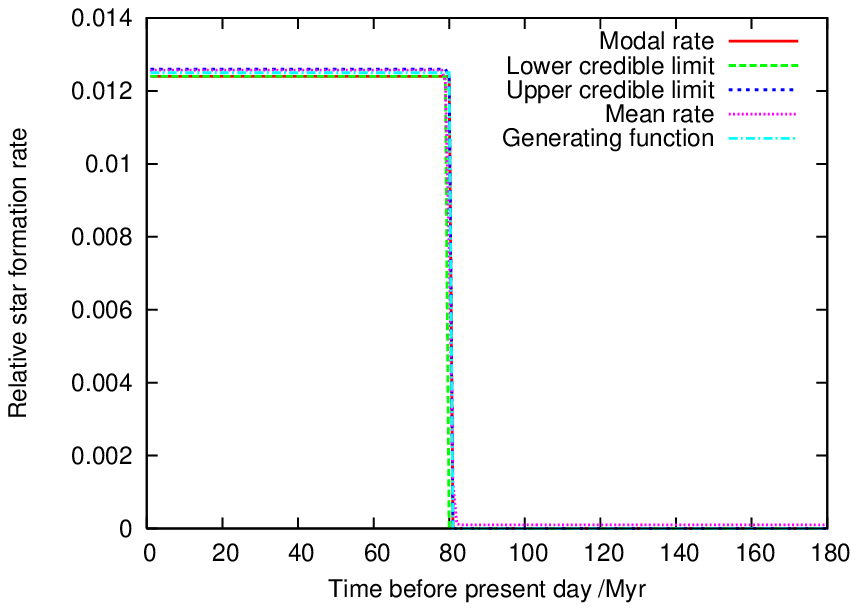}
\caption{The inferred star formation history of Population 4 with one rectangle. The errors are almost invisble.}
\label{fig:r22}
\includegraphics[width=0.5\textwidth]{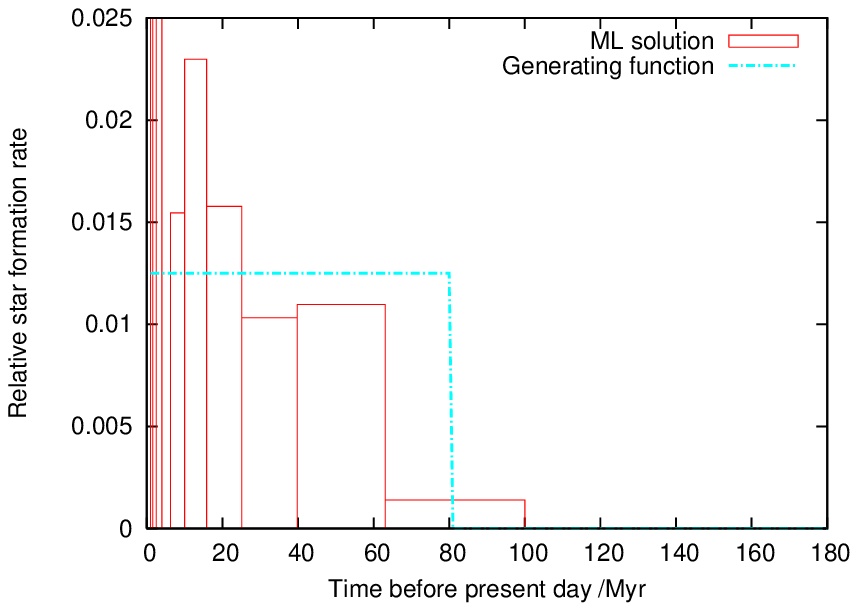}
\caption{The histogram-based maximum likelihood ($\mathcal{L}_2$) star formation history of Population 4.}
\label{fig:r4ml}
\end{figure}

 We generated 2,000 stars and found that three bursts gave the best fit
 in the Gaussian case but that the rate is a poor match
 (Fig.~\ref{fig:r2}), even with this abundance of data. The abrupt
 discontinuity at $80\, \rm Myr$ seems to have provoked a spike
 analogous to the Gibbs phenomenon found when a discontinuity is
 fitted with a Fourier series. By contrast one rectangle provides a
 far better match (Fig.~\ref{fig:r22}). The posterior probabilities of
 the different models are in favour of the rectangles by a ratio of
 over a thousand to one.

The histogram fit is shown in Fig.~\ref{fig:r4ml} and is surprisingly
poor. The logarithmic nature of the bins means that they become very
small for times more recent than 10 Myr, even though we increased the
bin size to 0.2 in the logarithm of time. This means that very small
intervals of star formation are being fitted. These intervals can only
be distinguished by the very small number of very rapidly-evolving
massive stars and so the potential for uncertainty is high.

\section{Application to data}

We applied our method to NGC 1313-F3-1, a young star cluster that
forms part of a set of seven for which \citet{2011A&A...532A.147L}
were able to recover resolved stellar populations. This intriguing
object shows evidence for two distinct bursts of recent star formation
and is an excellent test subject for our new method. We first
performed the analysis with $V-I$ against $V$ and with $B-V$ against
$V$. We then used the full data cube and compared our results and the
inferred parameters are listed in Table~\ref{tab:clusterA}.

\begin{figure} 
\centering
\includegraphics[width=0.5\textwidth]{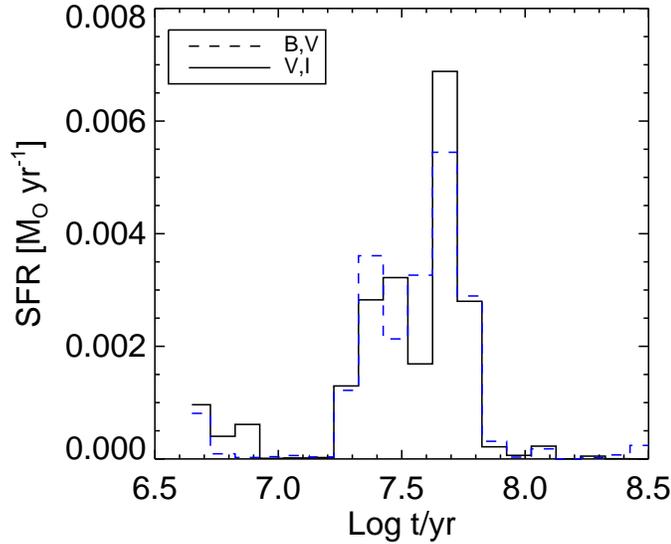}
\caption{The \citet{2011A&A...532A.147L} star formation history for NGC 1313-F3-1.}
\label{fig:lars}
\end{figure}

\citet{2011A&A...532A.147L} investigated the star formation history of
NGC 1313-F3-1 with the CMD-fitting code ${\rm \sc FITSFH}$
\citep{2010A&A...516A..10S}, which maximises the Poisson likelihood in
the same way as \citet{2002MNRAS.332...91D}. Their findings are shown in Fig.~\ref{fig:lars}.

\subsection{NGC 1313-F3-1 in two filters: V and V-I}

\begin{figure}
\centering
\includegraphics[width=0.5\textwidth]{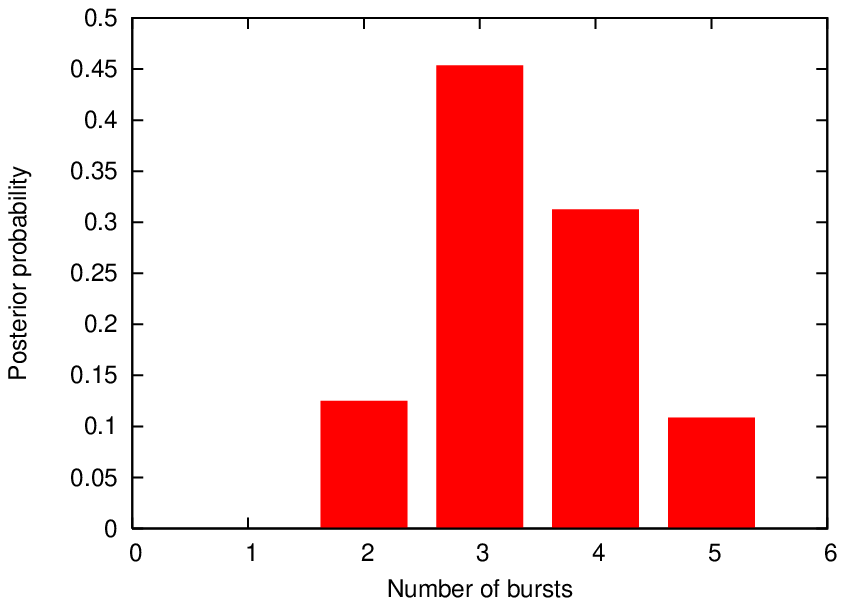}
\caption{The posterior probabilities of the number of bursts used to fit NGC 1313-F3-1 when $V$ and $V-I$ are used. Three bursts give the best fit.}
\label{fig:m3}
\includegraphics[width=0.5\textwidth]{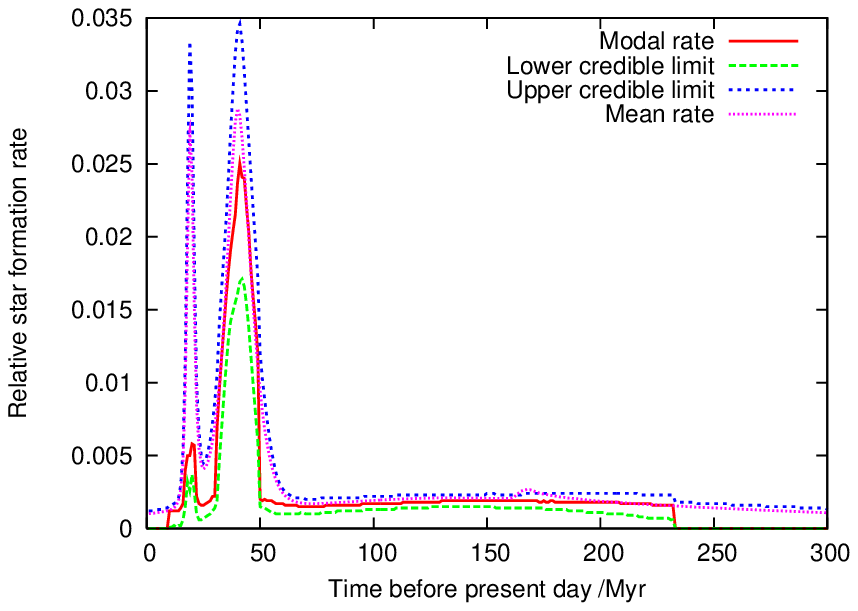}
\caption{The inferred star formation history of NGC 1313-F3-1 when $V$ and $V-I$ are used. The posterior probability of the star formation rate at each point in time is summarised by its mean, mode and minimum 68 per cent credible interval.}
\label{fig:r3}
\includegraphics[width=0.5\textwidth]{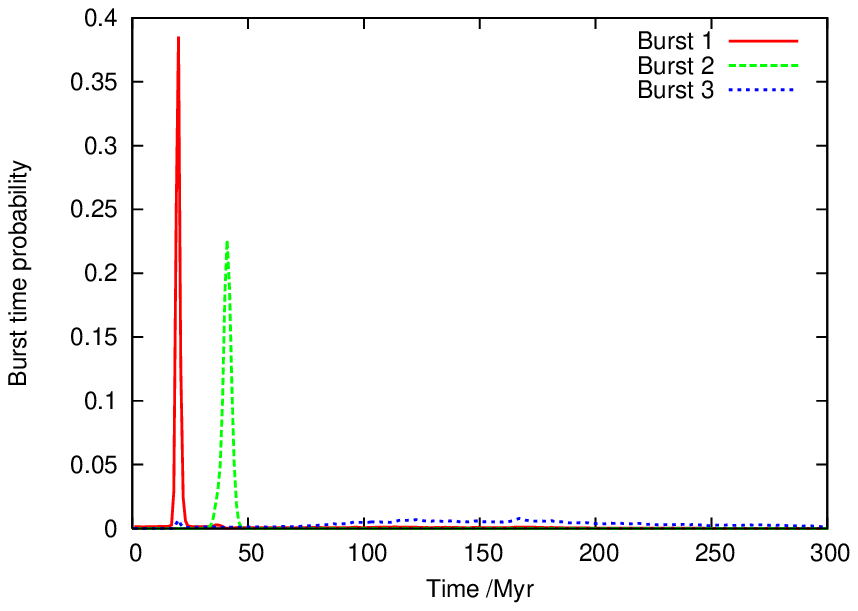}
\caption{The posterior probabilities of the times of NGC 1313-F3-1 with $V$ and $V-I$.}
\label{fig:t3}
\end{figure}

We applied a magnitude cut in $V$ at $-3$ and thus adopted the
completeness limits given in \citet{2011A&A...532A.147L}. This reduced
the number of stars under consideration to 122. We found that three
Gaussians provided the best fit (Fig.~\ref{fig:m3}) and the rate is
shown in Fig.~\ref{fig:r3}. Two of the Gaussians are distinct and
narrow bursts with well-defined times (Fig.~\ref{fig:t3}) but the
third is a very broad and ill-defined entity. It may represent a
continuous rate of low-level star formation; alternatively it could be
present to fit stars that, because of errors in the data and
inadequacies in the models, are not accounted for by the more recent
bursts.

\begin{figure} 
\centering
\includegraphics[width=0.5\textwidth]{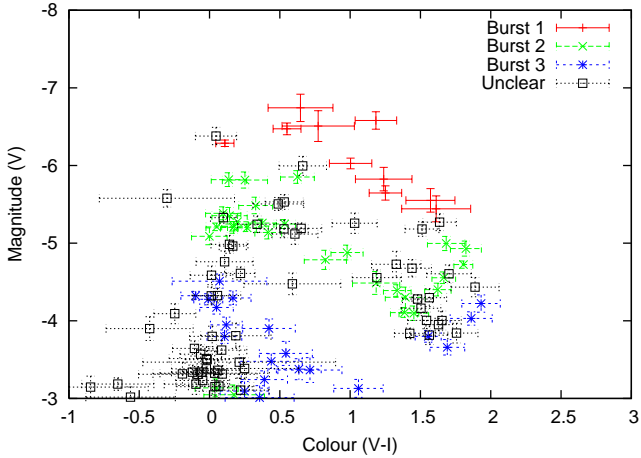}
\caption{The CMD with classified stars of NGC 1313-F3-1 with $V$ and $V-I$.}
\label{fig:c3}
\end{figure}

\begin{figure} 
\centering
\includegraphics[width=0.5\textwidth]{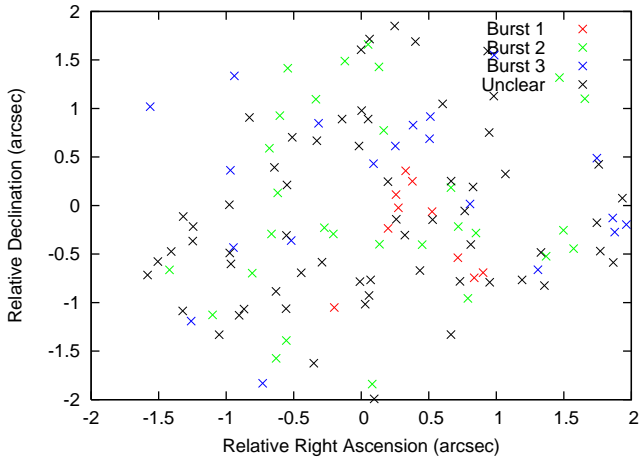}
\caption{The spatial locations of the classified stars of NGC 1313-F3-1 with $V$ and $V-I$. Those from Burst 1 appear to form a group.}
\label{fig:cc3}
\end{figure}

We found that the most recently-formed stars, as identified by the
classification probabilities in Fig.~\ref{fig:c3}, appear to form a
distinct group according to their spatial locations
(Fig.~\ref{fig:cc3}). This could be evidence for a distinct and
recently-formed population and is an unexpected bonus. Kinematic data
would help test this theory but unfortunately it is not available.

\begin{figure}
\centering
\includegraphics[width=0.5\textwidth]{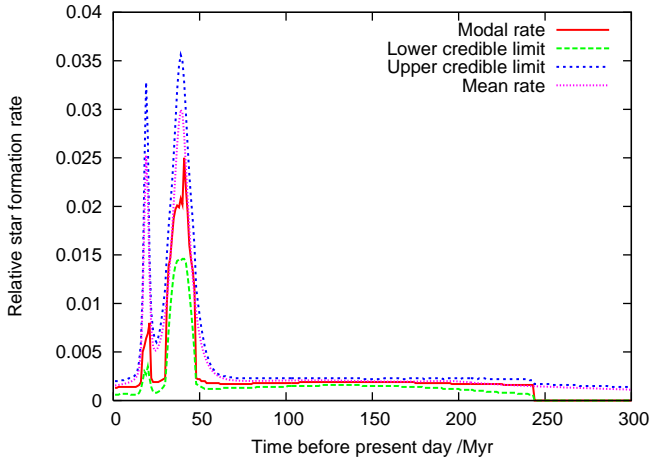}
\caption{The inferred star formation history of NGC 1313-F3-1 with $\mathcal{L}_1$ as the likelihood when $V$ and $V-I$ are used.}
\label{fig:r333}
\end{figure}

We repeated the analysis but with the likelihood calculated by the
Voronoi Tessellation error-blurring method ($\mathcal{L}_1$, from
Section 2.4).  The rate plot is shown in Fig.~\ref{fig:r333} and is
very similar to its counterpart (Fig.~\ref{fig:r3}).

\subsection{NGC 1313-F3-1 in two filters: V and B-V}

\begin{figure}
\centering
\includegraphics[width=0.5\textwidth]{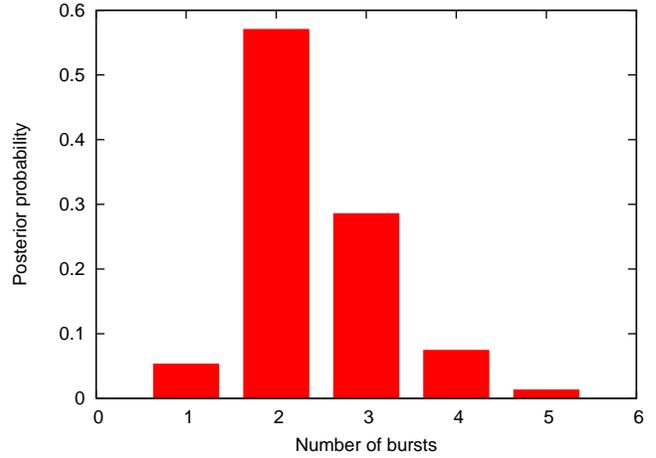}
\caption{The posterior probabilities of the number of bursts used to fit NGC 1313-F3-1 when $V$ and $B-V$ are used. Two bursts give the best fit.}
\label{fig:m3a}
\end{figure}

\begin{figure}
\centering
\includegraphics[width=0.5\textwidth]{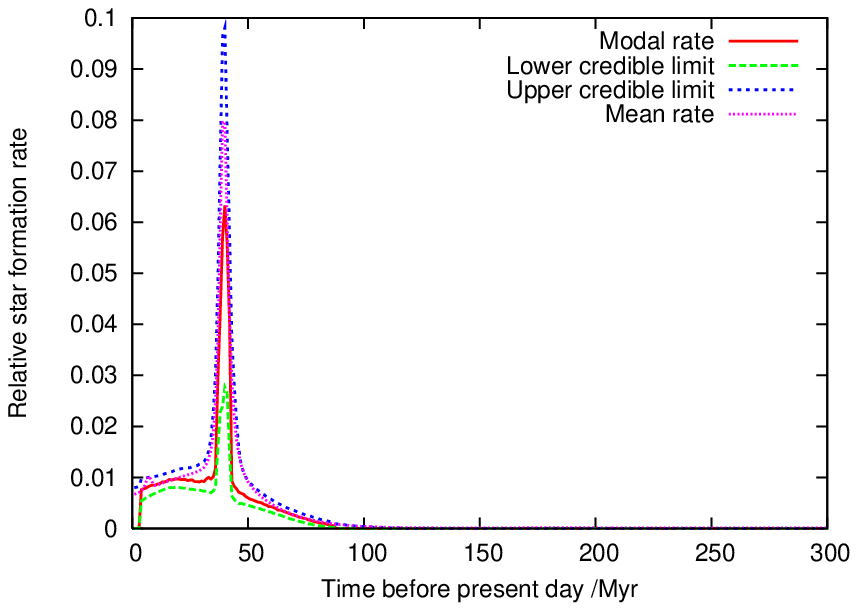}
\caption{The inferred star formation history of NGC 1313-F3-1 when $V$ and $B-V$ are used. The posterior probability of the star formation rate at each point in time is summarised by its mean, mode and minimum 68 per cent credible interval.}
\label{fig:r3a}
\end{figure}

\begin{figure}
\centering
\includegraphics[width=0.5\textwidth]{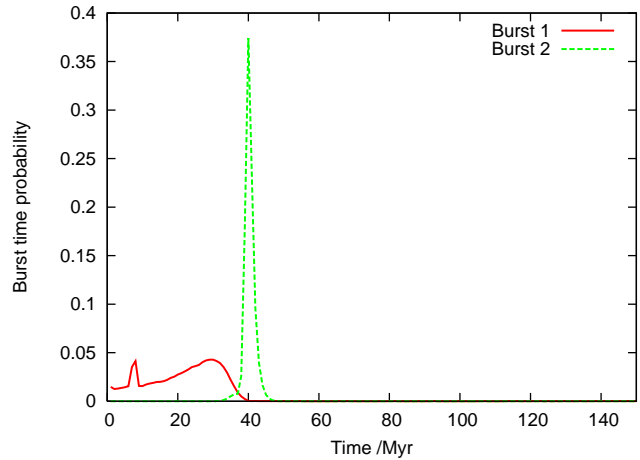}
\caption{The posterior probabilities of the times of NGC 1313-F3-1 with $V$ and $B-V$.}
\label{fig:t3a}
\end{figure}

We applied the same magnitude cut as before and on this occasion found
that two bursts provided the best fit (Fig.~\ref{fig:m3a}). The star formation history
is shown in Fig.~\ref{fig:r3a}. The earlier burst at $t_2=39^{+1}_{-1}$ Myr is
well defined (Fig.~\ref{fig:t3a}) and matches its counterpart from
the $V-I$ analysis. The most recent event is much more nebulously defined
and the very broad third burst is conspicuous by its absence.

\subsection{NGC 1313-F3-1 in three filters}

We finally construct the colour-colour-magnitude cube in $B-V$, $V-I$
and $V$ and compare with the data. We find that the most probable
number of Gaussians is two (Fig.~\ref{fig:m3a}) but that there is
reasonable posterior support for three bursts.

\begin{figure}
\centering
\includegraphics[width=0.5\textwidth]{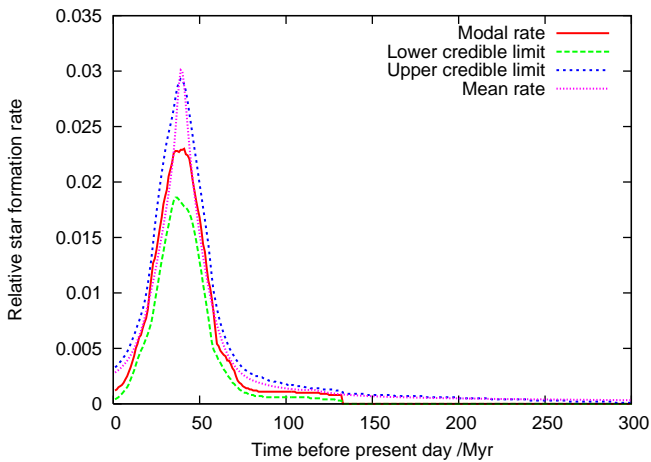}
\caption{The two burst inferred star formation history of NGC 1313-F3-1 when all the colours are used. The posterior probability of the star formation rate at each point in time is summarised by its mean, mode and minimum 68 per cent credible interval.}
\label{fig:r4}
\end{figure}

\begin{figure}
\centering
\includegraphics[width=0.5\textwidth]{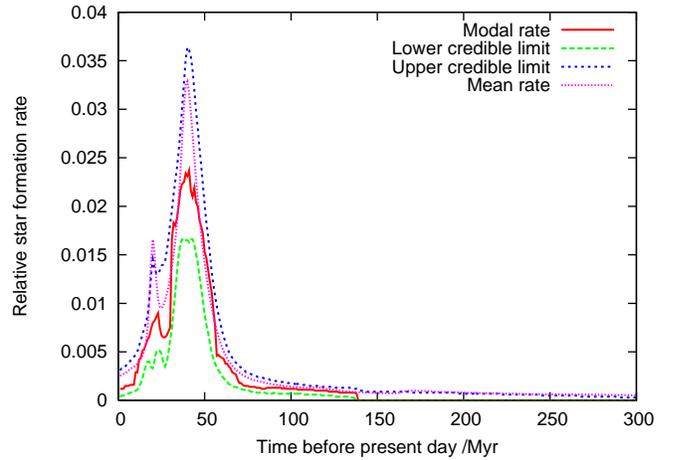}
\caption{The three burst inferred star formation history of NGC 1313-F3-1 when all the colours are used.}
\label{fig:r4a}
\end{figure}

\begin{figure}
\centering
\includegraphics[width=0.5\textwidth]{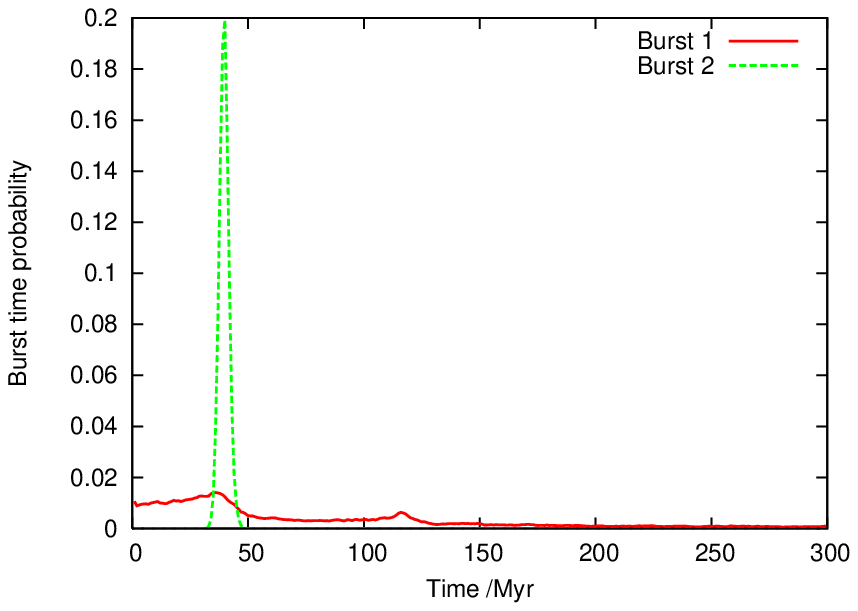}
\caption{The two burst posterior probabilities of the times of NGC 1313-F3-1 when all the colours are used.}
\label{fig:t4}
\end{figure}

\begin{figure}
\centering
\includegraphics[width=0.5\textwidth]{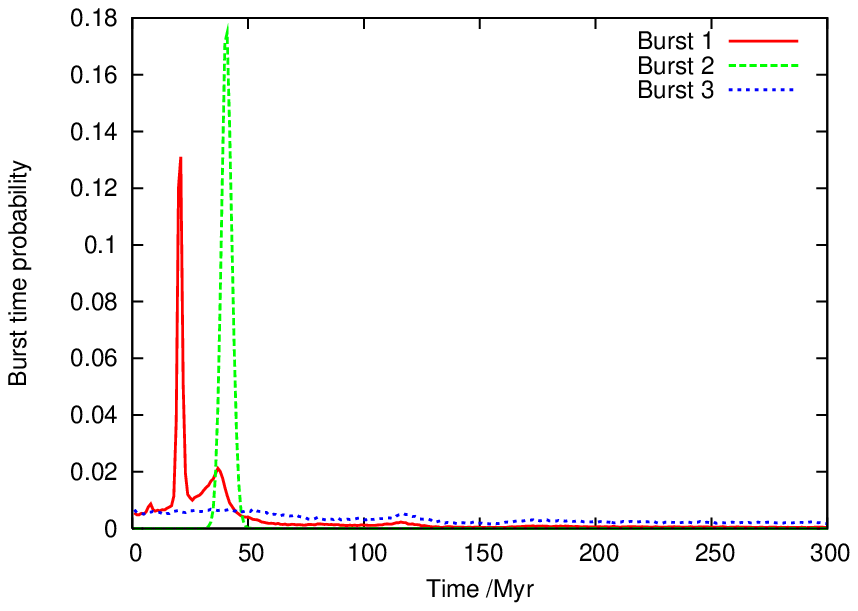}
\caption{The three burst posterior probabilities of the times of NGC 1313-F3-1 when all the colours are used.}
\label{fig:t4a}
\end{figure}

 We plot the rate with two (Fig.~\ref{fig:r4}) and three
 (Fig.~\ref{fig:r4a}) bursts and similarly the plot of the marginal
 probabilities of the times (Fig.~\ref{fig:t4} and
 Fig.~\ref{fig:t4a}). The earlier burst is amorphous enough that we
 may doubt its existence and ascribe it perhaps to uncertainties in
 the model. Alternatively such uncertainties could be the reason why
 the detection is not strong enough. This analysis demonstrates that
 to infer the star formation history of a resolved stellar population
 one must consider the probabilities of different scenarios rather
 than optimise some fitting function.

\begin{figure}
\centering
\includegraphics[width=0.5\textwidth]{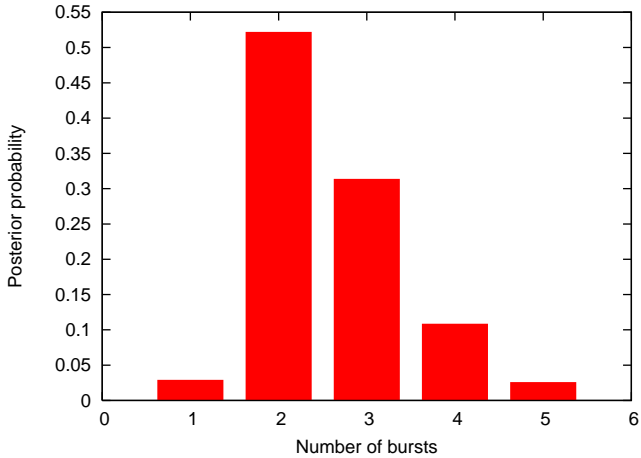}
\caption{The posterior probabilities of the number of bursts used to fit NGC 1313-F3-1. Two bursts give the best fit but there is reasonable support for three and so we consider both models.}
\label{fig:m4}
\end{figure}

\begin{table}
\caption{The estimated parameters of NGC 1313-F3-1.}
\begin{tabular}{| l | l | l | l |}
  \hline
   Burst & Time /Myr  & Width  /Myr & Mass fraction \\
  \hline
  \hline
 \multicolumn{4}{c}{$V$ and $V-I$ fitted with three bursts.}  \\
  \hline
  1 & $19^{+1}_{-1}$ & $1^{+1}_{-1}$ & $0.1^{+0.06}_{-0.03}$ \\
  2 & $40^{+2}_{-1}$ & $5^{+3}_{-2}$ & $0.39^{+0.1}_{-0.08}$\\
  3 & $166^{+54}_{-78}$ & $1^{+144}_{-1}$ & $0.47^{+0.11}_{-0.12}$\\
  \hline
 \multicolumn{4}{c}{$V$ and $B-V$ fitted with two bursts.}  \\
  \hline
  1 & $29^{+5}_{-16}$ & $1^{+32}_{-1}$ & $0.58^{+0.10}_{-0.10}$ \\
  2 & $39^{+1}_{-1}$ & $1^{+2}_{-1}$ & $0.41^{+0.10}_{-0.10}$\\
  \hline
 \multicolumn{4}{c}{$V$, $V-I$ and $B-V$ fitted with two bursts.}  \\
  \hline
  1 & $39^{+1}_{-2}$ & $13^{+3}_{-5}$ & $0.74^{+0.16}_{-0.21}$\\
  2 & $35^{+57}_{-35}$ & $1^{+120}_{-1}$ & $0.25^{+0.21}_{-0.16}$ \\
  \hline
 \multicolumn{4}{c}{$V$, $V-I$ and $B-V$ fitted with three bursts.}  \\
  \hline
  1 & $20^{+23}_{-7}$ & $1^{+31}_{-1}$ & $0.13^{+0.15}_{-0.13}$ \\
  2 & $40^{+2}_{-2}$ & $3^{+6}_{-2}$ & $0.50^{+0.18}_{-0.16}$\\
  3 & $451^{+111}_{-51}$ & $1^{+126}_{-1}$ & $0.25^{+0.10}_{-0.20}$\\
  \hline
\end{tabular}
\label{tab:clusterA}
\end{table}

\section{Conclusions}

We have demonstrated a new technique to infer the star formation
histories of stellar populations. Our Gaussian parametrisation is a
reasonable way of ensuring a continuous function of time. The use of
Reversal Jump MCMC means that we can automatically apply Occam's Razor
and find the number of Gaussians most appropriate for modelling the
data, thus avoiding overfitting. The Reversable Jump process also
allows the direct comparison of differently-parametrised models.

We note that it is possible to apply transdimensional Bayesian methods
to the traditional histogram parametrisation. The principal difficulty
would lie in the calculation of the Jacobian of the transformation
between $M$ time bins and, for example, $M+1$. We maintain though that
the Gaussian burst parametrisation is preferable. In the case when the
population really consists of distinct bursts then the advantages of
our method are obvious because we can infer the parameters of the
bursts. We can also infer the classification probabilities that a
given star was formed in a given burst.

Our method could also be easily adapted to fit the parameters used in
stellar models by making them into variables along with the star
formation history. With an appropriate prior one could then recover
the posterior probability distribution for the parameter given the
data under consideration. Naturally only very high quality data would
be suitable for this process and it would be wise to compare several
different data sets.

We are aware that we have been somewhat cavalier with the treatment of
errors and that a better approach would take into account
incompleteness as a function of magnitude, blending and model
uncertainties. The first of these could be addressed by the inclusion
of a parametrised form of the completeness function with appropriate
priors. A sigmoid between full and zero completeness with a location
parameter and a scale parameter is one possibility. These parameters
would then be varied in the MCMC process. One could similarly fit the
overall extinction and the distance modulus if these were
poorly-constrained.

More general improvements are also possible. The prior probabilities
for the masses and the widths could perhaps be improved by the
consideration of the behaviour of real star formation
bursts. Alternatively the priors could be parametrised and
hyperpriors placed on the parameters. An example of this would be to
allow $\alpha$, the concentration parameter of the symmetric Dirichlet
distribution, to vary in the MCMC process but with a hyperprior that
favoured $\alpha=1$ over $\alpha=100$ or $\alpha=0.01$. This approach
is probably a better way of representing our ignorance
\citep{stephens2}. The use of Gaussians is of course entirely
arbitrary and we are ready to believe that other parametrisations may
be more physically motivated and offer better fits. We compared
exponentials with Gaussians and found that the latter was favoured by
about three to one in the posterior probabilities; nevertheless there
may be other stellar populations where this is not true. In the future
it may be sensible to compare more general symmetric and asymmetric
functions.

\section*{Acknowledgements}
We thank the reviewer for many useful suggestions and Christopher
Berry for suggesting exponentials as an alternative
parametrisation. JJW was supported by a stipend from STFC, though
sadly this is no longer the case. Fig.~\ref{fig:lars} is from
\citet{2011A&A...532A.147L} and is reproduced with permission from
Astronomy \& Astrophysics, \textcopyright ESO.

\bibliographystyle{mn2e} \bibliography{CMD}

\end{document}